\documentclass[preprint,3p,twocolumn,letterpaper]{elsarticle}
\journal{Physics Letters B}
\usepackage{epsfig}
\usepackage[usenames,dvipsnames]{color}
\usepackage{amsmath}
\usepackage{graphicx}
\usepackage{amssymb}
\usepackage{hyperref}

\usepackage{widetext}

\usepackage{float}
\usepackage{bm}
\usepackage{dcolumn,multirow,subfig}

\usepackage{relsize}
\def\Babar{{\mbox{\slshape B\kern-0.1em{\smaller A}\kern-0.1em B\kern-0.1em{\smaller A\kern-0.2em R}}}}

\newcommand{\ba}{\begin{array}}
\newcommand{\ea}{\end{array}}
\def\beq{\begin{equation}}
\def\eeq{\end{equation}}
\def\bea{\begin{eqnarray}}
\def\eea{\end{eqnarray}}
\def\nn{\nonumber}

\def\roughly#1{\mathrel{\raise.3ex\hbox
{$#1$\kern-.75em\lower1ex\hbox{$\sim$}}}}

\def\gsim{\roughly>}
\def\sla#1{\raise.15ex\hbox{$/$}\kern-.57em #1}

\def\bd{B_d^0}

\def\order{\lower 1.8ex \hbox{\LARGE\~{}}}

\newcommand{\taurho}{\ensuremath{\tau^+\to\rho^+\nu\ }}
\newcommand{\taue}{\ensuremath{\tau^+\to e^+\nu\bar{\nu}\ }}
\newcommand{\taumu}{\ensuremath{\tau^+\to \mu^+\nu\bar{\nu}\ }}
\newcommand{\taupi}{\ensuremath{\tau^+\to\pi^+\nu\ }}

\def\bdtau{B\to D^{(\ast)}\tau\nu_{\tau}}

\def\bd0tau{B\to D \tau\nu_{\tau}}
\def\bdasttau{B\to D^{\ast}\tau\nu_{\tau}}

\def\be {\begin{equation}}
\def\ee {\end{equation}}

\begin{document}
  


\begin{frontmatter}

\title{Correlating new physics signals in $B \to D^{(*)} \tau \nu_{\tau}$ with $B \to \tau \nu_{\tau}$}

\author[IITG]{Soumitra Nandi}
\ead{soumitra.nandi@iitg.ernet.in}

\author[IITG]{Sunando K. Patra\corref{cor}}
\ead{sunando.patra@gmail.com}

\author[BNL]{Amarjit Soni}
\ead{adlersoni@gmail.com}

\cortext[cor]{Corresponding author}

\address[IITG]{Indian Institute of Technology, North Guwahati, Guwahati 781039, Assam, India }
\address[BNL]{Physics Department, Brookhaven National Laboratory, Upton, NY 11973, USA}

\begin{abstract}
Semileptonic and purely leptonic decays of B meson to $\tau$, such as $\bdtau$ and $B\to\tau\nu_\tau$ are studied. 
Recognizing that there already were some weak hints of possible deviations from the SM in the measurements of $\mathcal{B}(B\to\tau\nu_\tau)$ by \Babar~and Belle and the fact that detection of the $\tau$ also occurs in the measurements of $\bdtau$, we stress the importance of joint studies of these processes, whenever possible. 
For this purpose, as an illustration, we introduce the observable, $\mathcal{R}(D^{(*)})/\mathcal{B}(B\to\tau\nu_\tau)$ where, for one thing, the unknown systematics due to $\tau$ identification are expected to largely cancel. We show that all measurements of this observable are consistent with the existing data, within somewhat largish experimental errors, with the predictions of the SM. 
We stress that precise experimental measurement and comparison with theory of the branching ratio for $B \to \tau \nu_\tau$ is extremely important for a reliable search of new physics. 
Furthermore, in view of the anticipated improved precision in experiments in the next few years, in addition to $\mathcal{R}(D^{(*)})$, host of other ratios analogous to $\mathcal{R}(D^{(*)})/\mathcal{B}(B \to \tau \nu_{\tau})$ in the SM are suggested for lattice calculations as well, so that for more stringent tests of the SM, correlations in lattice calculations can be properly taken into account to enhance precision.
\end{abstract}


\end{frontmatter}

\section{Introduction}

\begin{figure}[hbt]
\centering
 \includegraphics[scale=0.4]{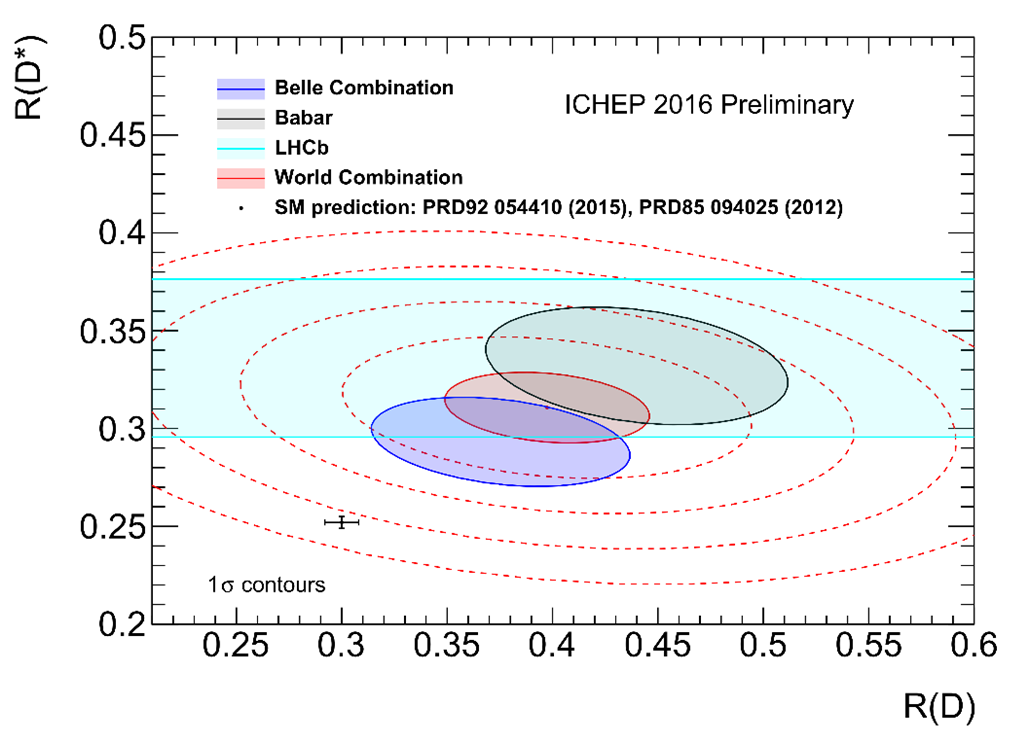}
 \caption{Current experimental status in the measurements of $R(D)$ and $R(D^*)$ \cite{belle_talk}.}
 \label{fig_hfag}
\end{figure}

The observed excess in the branching fractions of the semitaunic decays, $\bd0tau$ and $\bdasttau$, 
has drawn a lot of attention in the recent years. 
The present experimental status is summarized in Fig. \ref{fig_hfag} \cite{belle_talk}. ,

Here, $R(D)$ and $R(D^*)$ are defined as 

\begin{align}
\nn \mathcal{R}(D) &= \frac{\mathcal{B}\left(\overline{B} \to D \tau^- \overline{\nu}_{\tau}\right)}
{\mathcal{B}\left(\overline{B} \to D l^- \overline{\nu}_{l}\right)}\,, \\
\mathcal{R}(D^*) &= 
\frac{\mathcal{B}\left(\overline{B} \to D^* \tau^- \overline{\nu}_{\tau}\right)}{\mathcal{B}
\left(\overline{B} \to D^* l^- \overline{\nu}_{l}\right)}\,.
\end{align}

As had been emphasized  in several works \cite{Kiers:1997zt,Chen:2006nua,Kamenik:2008tj,Nierste:2008qe,Fajfer:2012vx,
Na:2015kha}, the theory uncertainties in these observables are only a few percent, being   
independent of the CKM element $|V_{cb}|$ and also to a large extent, of the form-factors~ 
\footnote{We note here the extremely small (SM) theory error quoted in $R(D^*)$(a lot smaller than that of $R(D)$) and emphasize that 
so far no lattice calculation of the ratio involving $D^*$ over the full kinematic region exists. Fermilab calculation 
\cite{Bailey:2014tva} is only at the end point.}.
Interestingly, the $\mathcal{R}(D^{(*)})$ values measured by \Babar~\cite{Lees:2013uzd}   
exceed SM expectations by $2.0 \sigma$ and as much as  $2.7 \sigma$ respectively, 
and if taken together, disagree with the SM by about $3.4 \sigma$.
On the other hand, older Belle results used to lie in between the SM expectation and the 
\Babar~measurement and were consistent with both \cite{Huschle:2015rga,Abdesselam:2016cgx}. 
Latest Belle results \cite{Abdesselam:2016xqt} using the full data sample of $772 \times 10^6$ $B \bar{B}$ pairs is 
completely consistent with SM within $0.6~\sigma$. 
LHCb announced the results of their first measurement of $\mathcal{R}(D^{*})$ \cite{Aaij:2015yra}, and their 
result is $2.1 \sigma$ larger than the value expected in SM. Both \Babar~and Belle have analyzed the effects 
of the charged Higgs of type 2HDM-II on $\mathcal{R}(D^{(*)})$. \Babar~analysis shows that in 2HDM-II, the measured values 
of $\mathcal{R}(D)$ and $\mathcal{R}(D^{*})$ can not be explained simultaneously by any 
point in the $\tan{\beta}$ - $m_{H^+}$ parameter space which is allowed by the data.
Hence, 2HDM-II as a possible new physics (NP) candidate was excluded by \Babar~at $99.8\%$ confidence level, 
while the earlier Belle measurement is consistent with the 2HDM-II prediction given in the regions around 
$\tan{\beta}/m_{H^+} = 0.45\text{GeV}$ \cite{Huschle:2015rga}. As is already mentioned, latest Belle result \cite{Abdesselam:2016xqt}
does not require 2HDM or any other NP. Also, since LHCb cannot do $B \to D$ semileptonic decays, they cannot make statements about 
2HDM-II.

In passing we mention that a similar pattern, but with even bigger errors, is observed in the \Babar~ and Belle measurements
on the branching fractions of the purely leptonic decay $B^+ \to \tau^+ \nu_{\tau}$ which uses both the 
leptonic and hadronic channels for the identification of $\tau$. 
Note in particular that, both the experiments had earlier used only the leptonic channels for the identification of $\tau$ in the measurements of $\mathcal{R(D^{(*)})}$. Only in their latest analysis \cite{Abdesselam:2016xqt}, Belle has used hadronic channels for $\tau$ reconstruction.

We think it is rather useful  to examine $\mathcal{B}(B \to \tau \nu)$ simultaneously, when possible, 
with $R(D^{(*)})$ for a variety of reasons. For one, if $R(D^{(*)})$ is showing deviation(s) from the SM, then 
it stands to reason that $\mathcal{B}(B \to \tau \nu)$ may also do the same. Also $R(D^{(*)})$ may be suffering from background 
contaminations (e.g. from higher charm resonances) that are difficult to deal with, whereas $\mathcal{B}(B \to \tau \nu)$ 
may not have that difficulty and therefore to that extent may be more reliable. 

Moreover, since detection of $\tau$ is involved both in $R(D^{(*)})$ and in $\mathcal{B}(B \to \tau \nu)$, by studying them together 
as by our  proposed ratio, eq.(\ref{rtaudeq}), if there are any unknown systematics affecting the $\tau$ detection then they will 
tend to largely cancel. Finally, in a large class of new physics models that affect the $\tau - \nu$ vertex, the effect of 
new physics will tend to cancel in the ratio $R_{\tau}(D^{(*)})$ so this ratio may serve as a very good diagnostic of the new physics.

Of course, the experimental measurement of $\mathcal{B}(B \to \tau \nu)$ in itself is extremely demanding and the accuracy of the 
measurements to date are fairly limited (as elaborated in section \ref{sec:smexp} below) but our point in suggesting these correlation(s) is 
to emphasize their importance for the long run; in particular, in view of the much larger data sets that will become available 
from Belle-II in the near future.


Moreover, since $\mathcal{R}(D)$ and $\mathcal{R}(D^*)$ are independent of $|V_{cb}|$, we emphasize the importance of analogous
ratios for semileptonic decays $B \to \pi (\rho,\omega)~\ell(\tau)~\nu$ and similarly for $B \to D^{(*)} ~\ell(\tau) ~\nu$ (see eq.(\ref{rtaupi})).

\section{SM vs. Experiments}\label{sec:smexp}
\subsection{$\mathcal{B}(B^+ \rightarrow \tau^+ \nu_{\tau})$}

In Tables \ref{tab:tauBABAR} and \ref{tab:tauBelle}, the \Babar~and Belle measured values of the
$Br(B\to \tau\nu_\tau)$ are shown, using leptonic and hadronic decays of $\tau$ separately. 
Their combined results are also shown, and they are consistent with each other within errors.   

\begin{table}[hbt]
\begin{center}
\resizebox{0.5\textwidth}{!}{%
\begin{tabular}{lccc}
\hline
\hline
Decay Mode    &  $\epsilon_k (\times 10^{-4})$ & Signal yield & $\mathcal{B} (\times 10^{-4})$ \\
\hline
\noalign{\vskip1pt}
\taue  &   $2.47\pm0.14$     & $4.1 \pm 9.1$ & $0.35 ^{+0.84}_{-0.73}$ \\
\taumu &  $2.45\pm0.14$    & $12.9 \pm 9.7$  & $1.12 ^{+0.90}_{-0.78}$ \\
\noalign{\vskip1pt}
\hline
\noalign{\vskip1pt}
Leptonic & $4.92 \pm 0.198$ & $17 \pm 13.3$ & $0.739 \pm 0.578$ \\
\noalign{\vskip1pt}
\hline
\noalign{\vskip1pt}
\taupi &  $0.98\pm0.14$      & $17.1 \pm 6.2$ & $3.69   ^{+1.42}_{-1.22}$ \\
\taurho & $1.35\pm0.11$     & $24.0 \pm 10.0 $ & $3.78   ^{+1.65}_{-1.45}$ \\
\noalign{\vskip1pt}
\hline
\noalign{\vskip1pt}
Hadronic & $2.33 \pm 0.178$ & $41.1 \pm 11.77$ & $3.77 \pm 1.12$ \\
\noalign{\vskip1pt}
\hline
\noalign{\vskip1pt}
\hline
\noalign{\vskip1pt}
combined &  & $ 62.1 \pm 17.3$ & $1.83^{+0.53}_{-0.49}$ \\ 
\hline
\noalign{\vskip1pt}
\hline
\end{tabular}
}
\caption{{The measured values of $\mathcal{B}(B^+\to\tau\nu_{\tau})$ by \Babar~in various $\tau$ decay modes, and their 
combined value for $N_{B \bar{B}} = (467.8 \pm 5.1) \times 10^6$ \cite{Lees:2012ju}.}}
\label{tab:tauBABAR}
\end{center}
\end{table}
\begin{table}[hbt]
\begin{center}
\resizebox{0.5\textwidth}{!}{%
\begin{tabular}{lccc}
\hline
\hline
Decay Mode    &  $\epsilon_k (\times 10^{-4})$ & Signal yield & $\mathcal{B} (\times 10^{-4})$ \\
\hline
\noalign{\vskip1pt}
$\tau^+ \to e^+ \bar{\nu_{\tau}}\nu_{e}$  &   $6.8$     & $47 \pm 25$ & $0.90 \pm 0.47$ \\
$\tau^+ \to \mu^+ \bar{\nu_{\tau}}\nu_{\mu}$ &  $5.1$    & $13 \pm 21$  & $0.34 \pm 0.55$ \\
\noalign{\vskip1pt}
\hline
\noalign{\vskip1pt}
Leptonic & 11.9 & $60 \pm 32.65$ & $0.653 \pm 0.355$ \\
\noalign{\vskip1pt}
\hline
\noalign{\vskip1pt}
$\tau^+ \to \pi^+ \bar{\nu_{\tau}}$ &  $4.0$      & $57 \pm 21$ & $1.82 \pm 0.68$ \\
$\tau^+ \to \pi^+ \pi^0 \bar{\nu_{\tau}}$ & $7.2$     & $119 \pm 33$ & $2.16 \pm 0.60$ \\
\noalign{\vskip1pt}
\hline
\noalign{\vskip1pt}
Hadronic & 11.2 & $176 \pm 39.12$ & $2.036 \pm 0.452$ \\
\noalign{\vskip1pt}
\hline
\noalign{\vskip1pt}
\hline
\noalign{\vskip1pt}
combined & 23.1 & $222 \pm 50$ & $1.25 \pm 0.28$ \\ 
\hline
\noalign{\vskip1pt}
\hline
\end{tabular}
}
\caption{The measured values of $\mathcal{B}(B^+\to\tau\nu_{\tau})$ by Belle in various $\tau$ decay modes, and their 
combined value for $N_{B^+ B^-} = 772 \times 10^6$ \cite{Kronenbitter:2015kls}.}
\label{tab:tauBelle}
\end{center}
\end{table}

The expression for the branching fraction $Br(B\to \tau\nu_\tau)$ in the SM is given by  
\begin{align}\label{brb2tn}
 \nn \mathcal{B}_{SM}&(B^+ \rightarrow \tau^+ \nu_{\tau}) \\
 &= \frac{G^2_F m_B m^2_{\tau}}{8 \pi} \left[1 - \frac{m^2_{\tau}}{m^2_B}\right]^2 f^2_B \left|V_{ub}\right|^2 \tau_{B^+},
\end{align}
where $G_F$ is the Fermi constant, $m_B$ and $m_{\tau}$, are the $B^+$ meson and $\tau$ lepton masses respectively, 
$\tau_{B^+}$ is the $B^+$ meson lifetime. The branching fraction is sensitive to the 
$B$ meson decay constant $f_B$ and the Cabibbo-Kobayashi-Maskawa (CKM) matrix element $|V_{ub}|$.
With the numerical values of all the relevant parameters listed in table \ref{tab:params}, we obtain
\be
\mathcal{B}_{SM}(B^+ \rightarrow \tau^+ \nu_{\tau}) = (0.947 \pm 0.182)\times 10^{-4}.
\ee
Numerical value of the CKM element $|V_{ub}|$ is obtained after fitting latest lattice calculation of 
$B \to \pi l \nu$ form factors with the experimental measurements of the branching fraction from 
\Babar~and Belle, leaving the relative normalization as a free parameter, for details see Ref. \cite{Flynn:2015mha,Lattice:2015tia}. 
Here, we use $|V_{ub}|$ i.e $|V_{ub}^{Ex}|$ with a more conservative error from \cite{Flynn:2015mha} than that of HFAG or PDG (see table 
\ref{tab:params}) or from \cite{Lattice:2015tia}). Part of the reason we now feel more confident about this exclusive value of $V_{ub}$ is that 
it is found to be in excellent agreement with the value determined from exclusive baryonic B-decays~\cite{Detmold:2015aaa}.

\begin{figure}[hbt]
\centering
 \includegraphics[scale=0.4]{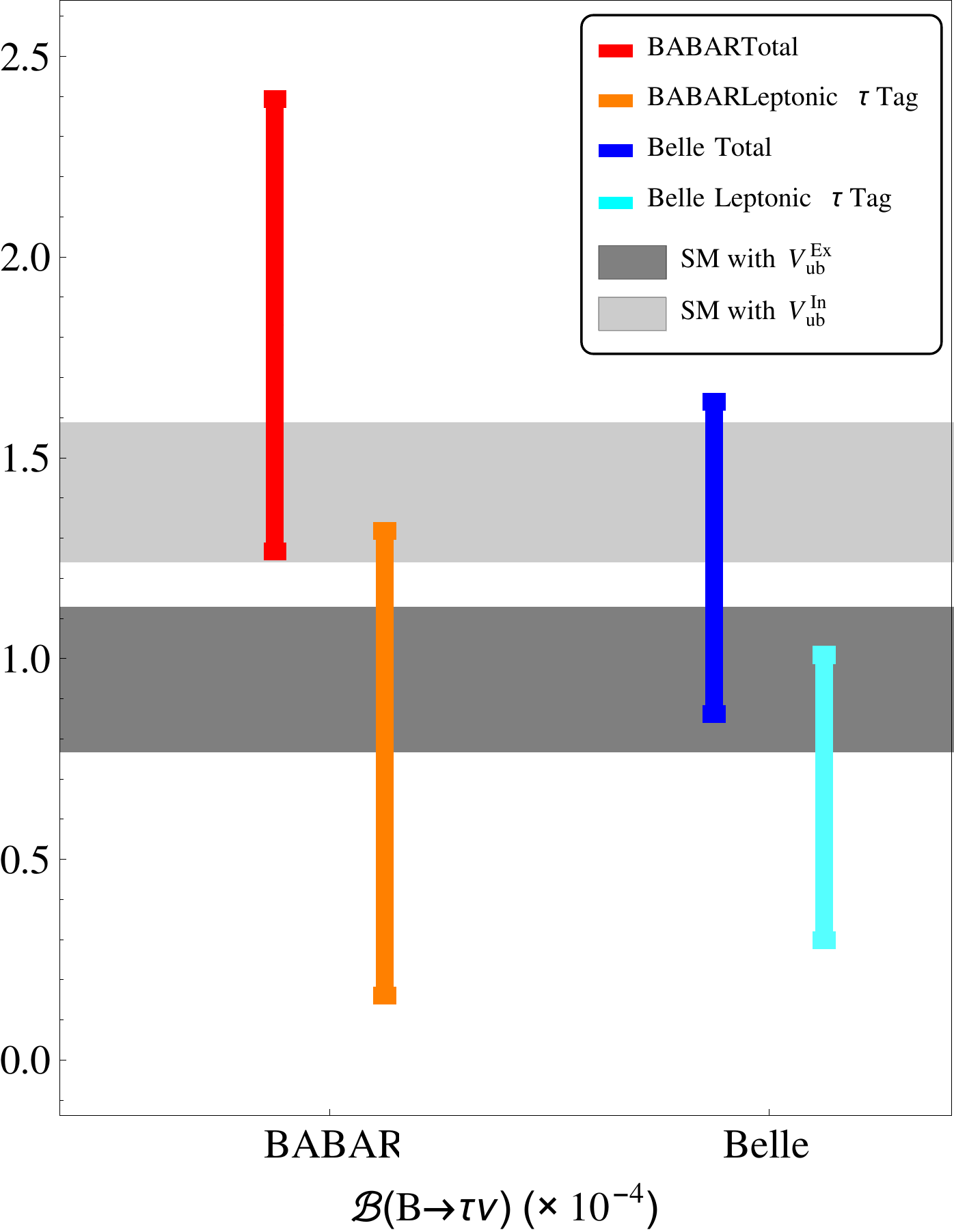}
 \caption{Graphical representation of the data shown in tables \ref{tab:tauBABAR}) and (\ref{tab:tauBelle}).}
\label{btn}
\end{figure}

In Fig. \ref{btn} the experimental measurements on $Br(B\to \tau\nu_\tau)$ are compared with the  
SM predictions. In this figure, for the sake of completion, we also  show the estimated branching fraction using the inclusive measurement of $|V_{ub}|$, 
i.e $|V_{ub}^{In}|$ (table \ref{tab:params}). The corresponding 
value of the branching fraction is given by. 
\be
\mathcal{B}_{SM}^{in}(B^+ \rightarrow \tau^+ \nu_{\tau}) = (1.413 \pm 0.175)\times 10^{-4}. 
\ee
From Fig. \ref{btn}, we see that Belle measurement is roughly consistent with the SM irrespective of which 
$V_{ub}$ (inclusive or exclusive) is used and \Babar~measurement just mildly disfavors the SM when 
exclusive $V_{ub}$ is used. On the other hand, if we consider only modes with leptonically reconstructed $\tau$s, \Babar~is consistent with both $V_{ub}$ values and Belle measurement slightly disfavors the inclusive $V_{ub}$.

\begin{table}[hbt]
\begin{center}
\begin{tabular}{lccc}
\hline
\hline
Parameters 	& Values\\
\hline
\noalign{\vskip1pt}
$f_B$ 		 & $0.191 \pm 0.007$ GeV \cite{Na:2012kp,Aoki:2013ldr}\\
$G_F$ 		 & $1.1663787(6) \times 10^{-5}$ GeV$^{-2}$ \cite{GF}\\
$m_B$ 		 & $5.27929 \pm 0.00015$ GeV \cite{mB} \\
$m_{\tau}$ 	 & $1.77686 \pm 0.00012$ GeV \cite{mtau} \\
$\tau_{B^+}$	 & $1.638(4)$ ps$^{-1}$ \cite{tauB+} \\
$\tau_{B^0}$	 & $1.520(4)$ ps$^{-1}$ \cite{tauB+} \\
$|V_{ub}^{Ex}|$  	 & $(3.61 \pm 0.32) \times 10^{-3}$  \cite{Flynn:2015mha} \\
$|V_{ub}^{In}|$  & $(4.41 \pm 0.15 ~^{+0.15}_{-0.19}) \times 10^{-3}$  \cite{Amsler:2008zzb} \\
$|V_{cb}|$ 	 & $(42.21 \pm 0.78) \times 10^{-3}$  \cite{Alberti:2014yda} \\
$m_b(\mu = m_b)$ & $4.20 \pm 0.07$  GeV\cite{Xing:2007fb}\\
$m_c(\mu = m_b)$ & $0.901^{+ 0.111}_{- 0.113}$ GeV\cite{Xing:2007fb}\\
$m_u$ 		 & $0.00236 \pm 0.00024$ GeV\cite{Carrasco:2014cwa}\\
$\lambda_1$	 & $-0.15 \pm 0.15$ \cite{Grossman:1994ax}\\
$\lambda_2$	 & $0.12 \pm 0.01$ \cite{Grossman:1994ax}\\
\hline
\noalign{\vskip1pt}
\hline
\end{tabular}
\caption{Input parameters used in obtaining theory predictions.}
\label{tab:params}
\end{center}
\end{table}


\begin{figure*}[hbt]
\centering
\includegraphics[scale=0.45]{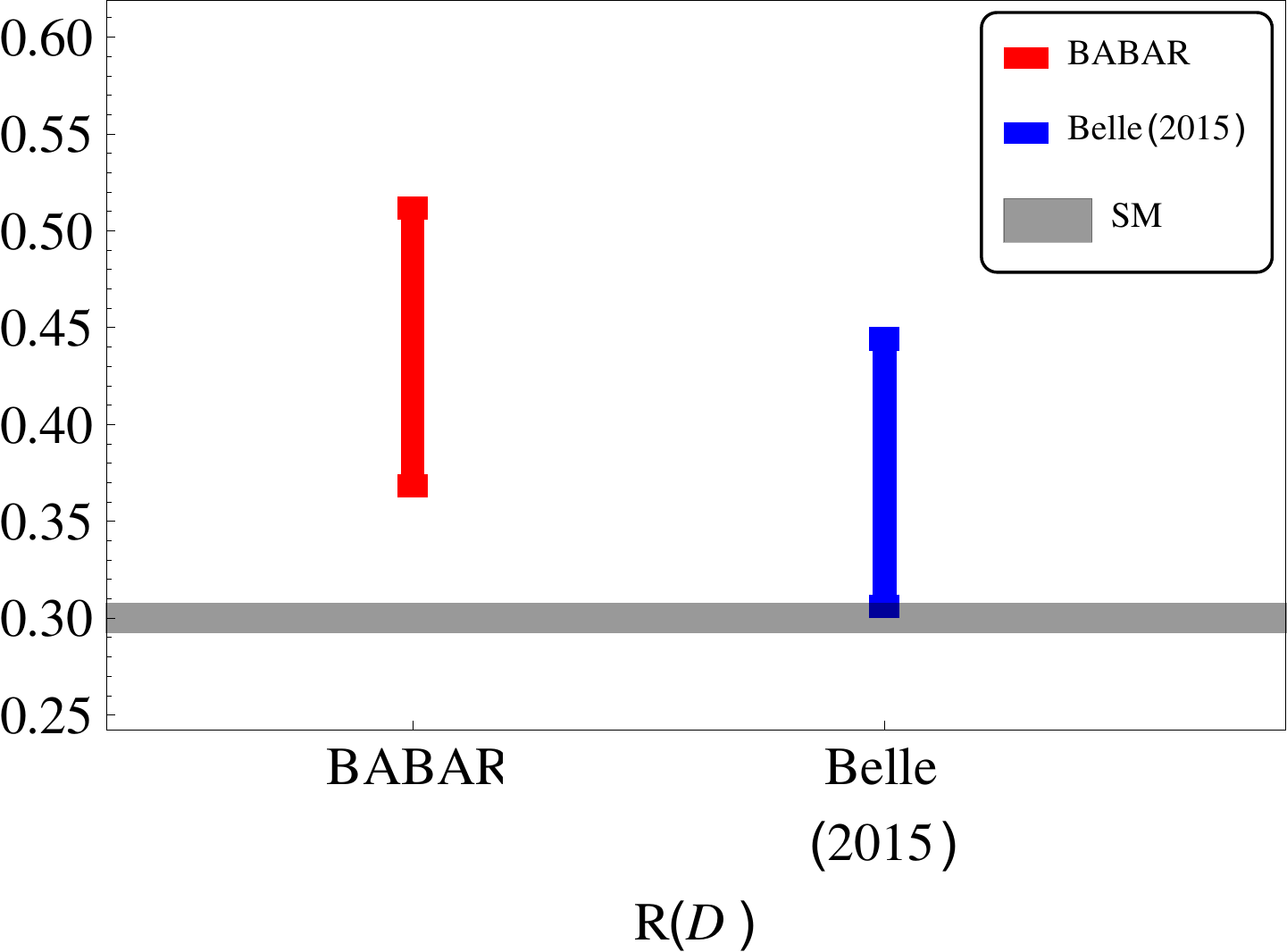}~~\includegraphics[scale=0.45]{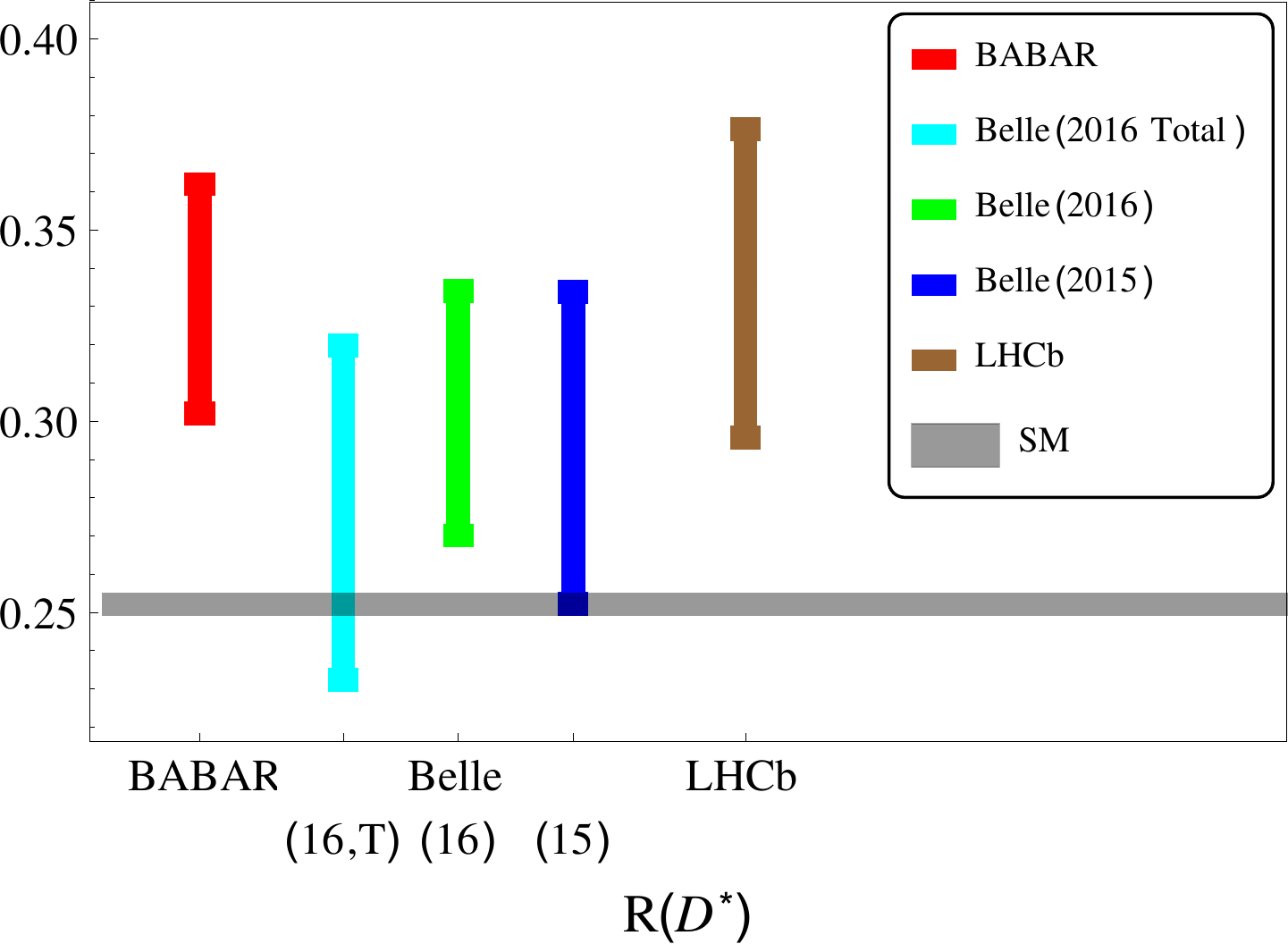}
\caption{Graphical representation of the data given in Table \ref{tab:RDst}. The Dark gray SM bands use the data from the first row of the table. The three vertical lines for Belle data of $R(D^*)$ are, respectively, (from right to left) Belle(2015)\cite{Huschle:2015rga}(dark blue), Belle(2016)\cite{Abdesselam:2016cgx}(green) and latest Belle data(2016) with full dataset\cite{Abdesselam:2016xqt}(cyan).}
 \label{rdrdst}
\end{figure*}

\subsection{$\mathcal{R}(D^{(*)})$ and $R_{\tau}(D^{(*)})$ }

For the measurements of branching fractions $Br(\bdtau)$, both \Babar~\cite{Lees:2013uzd} and Belle \cite{Huschle:2015rga}
use purely hadronic  tagging of ``the other B" and purely leptonic $\tau$ decays coming from the B undergoing semi-tauonic
decay. In one of their analyses(in 2016), Belle had used semi-leptonic tagging of ``the other B'' and measured $R(D^*)$ \cite{Abdesselam:2016cgx}. They are yet to publish their result on $R(D)$ using the same semileptonic tagging method for the 
other B. 
Most recently, Belle has published another result on $R(D^*)$ along with their first measurement of $P_{\tau}$ \cite{Abdesselam:2016xqt} with the total available dataset of 772 million $B \bar{B}$ pairs. In that analysis, they have 
used hadronic $\tau$ decays for $\tau$ reconstruction and hadronic tagging for the other $B$ ($B_{tag}$). In table (\ref{tab:RDst}) we list those measured values of $\mathcal{R}(D)$ and $\mathcal{R}(D^{*})$ along with their SM predictions. The same data has been plotted in Fig.\ref{rdrdst}. We note that like 
$\mathcal{B}(B\to \tau\nu_{\tau})$ the Belle 2015 data is consistent with the SM prediction, while 
the \Babar~data exceeds the SM expectations by $2.0 \sigma$ and as much as $2.7 \sigma$ respectively for 
$\mathcal{R}(D)$ and $\mathcal{R}(D^*)$. On the other hand, while Belle measurement of $R(D^*)$(2016) with leptonically tagged $\tau$s\cite{Abdesselam:2016cgx} is away from the corresponding SM prediction by $1.6 \sigma$, the most recent measurement with the full dataset is consistent with SM within $0.6 \sigma$ \cite{Abdesselam:2016xqt}.

\begin{table}[bt]
\begin{center}
\resizebox{0.5\textwidth}{!}{%
\renewcommand{\arraystretch}{1.2}
\begin{tabular}{lll}
\hline
\hline
	& $\mathcal{R}(D)$  			& $\mathcal{R}(D^*)$ \\
\hline
\noalign{\vskip1pt}
SM  	& $0.300 \pm 0.008$ \cite{Na:2015kha} &  $0.252 \pm 0.003$ \cite{Kamenik:2008tj} \\
 & $0.299 \pm 0.011$ \cite{Lattice:2015rga} & $~~$- \\
 & $0.299 \pm 0.003$ \cite{Bigi:2016mdz} & $~~$- \\
\Babar~   & $0.440 \pm 0.058 \pm 0.042 $ & $0.332 \pm 0.024 \pm 0.018 $ \cite{Lees:2013uzd}\\
Belle (2015)  & $0.375 \pm 0.064 \pm 0.026 $ & $0.293 \pm 0.038 \pm 0.015 $ \cite{Huschle:2015rga}\\
Belle (2016)  & -& $0.302 \pm 0.030 \pm 0.011 $ \cite{Abdesselam:2016cgx}\\
Belle(2016, Tot. Data.)  & -& $0.276 \pm 0.034 ^{+ 0.029}_{- 0.026}$ \cite{Abdesselam:2016xqt}\\
LHCb   	& - & $0.336 \pm 0.027 \pm 0.030 $ \cite{Aaij:2015yra}\\
\hline
\hline
\end{tabular}
}
\caption{The SM predictions and the experimentally measured values of $\mathcal{R}(D^{(\ast)})$. For experimental results, 
the first uncertainty is statistical and the second one is systematic. The third row represents the Belle results published 
in 2015 \cite{Huschle:2015rga}. The fourth row represents Belle's 2016 results of $R(D^*)$ \cite{Abdesselam:2016cgx}, wherein the tagging method for ``the other B'' is different from the previous analysis. The fifth row represents Belle's most recent results of $R(D^*)$ \cite{Abdesselam:2016xqt} with total available dataset.}
\label{tab:RDst}
\end{center}
\end{table}
\begin{table}[bt]
\begin{center}
\resizebox{0.5\textwidth}{!}{%
\renewcommand{\arraystretch}{1.2}
\begin{tabular}{lll}
\hline
\hline
	& $R_{\tau}(D)$ ($\times 10^3$)	& $R_{\tau}(D^*)$ ($\times 10^3$) \\
\hline
\noalign{\vskip1pt}
SM (With $V_{ub}^{Ex}$) & $3.17 \pm 0.61$		& $2.66 \pm 0.51$ \\
SM (With $V_{ub}^{In}$) & $2.12 \pm 0.27$		& $1.78 \pm 0.22$ \\
\Babar(Leptonic $\tau$ Tag)	   	& $5.96 \pm 2.26$		& $4.49 \pm 3.54$ \\
Belle(2015, Leptonic Tag)	   	& $5.7 \pm 3.3$			& $4.49 \pm 3.54$ \\
Belle(2016, Leptonic Tag)   &        -          &    $4.62 \pm 2.56$              \\
Belle(2016, Total Dataset,   &        -          &    $1.36 \pm 0.37$              \\
 Hadronic Tag)  &  &  \\
\hline
\hline
\end{tabular}
}
\caption{Our estimated values of $R_{\tau}(D^{(*)})$ using \Babar~and Belle measured values of $\mathcal{R}(D^{(*)})$.
The values in the last four rows are obtained using corresponding results on $R(D^{(*)}$ listed in table \ref{tab:RDst}. For the SM value, we use the first row of \ref{tab:RDst}.
}
\label{tab:RtauDst}
\end{center}
\end{table}

As mentioned in the introduction, 
to study the possibility of correlation in $\tau$ decays affecting the analyses and also for other potentially useful purposes, here we define a new observable 
$R_{\tau}(D^{(*)})$ as
\be
R_{\tau}(D^{(*)}) = \frac{\mathcal{R}(D^{(*)})}{\mathcal{B}(B^+ \rightarrow \tau^+ \nu_{\tau})}
\label{rtaudeq}
\ee
where $\mathcal{R}(D^{(*)})$ is normalized by $\mathcal{B}(B^+ \rightarrow \tau^+ \nu_{\tau})$. 

In order to explicitly spell out the possible cancellation of $\tau$ systematics in this ratio, 
$R_{\tau}(D^{(*)})$ can be defined as explained below.

The definition of $\mathcal{R}(D^{(*)})$ as used in the experimental analyses is the average of all the $\mathcal{R}(D^{(*)})^i$,
 which are given as
\be
\mathcal{R}(D^{(*)})^i = \frac{1}{{\cal B}^i_{\tau}} \frac{N_{sig} \epsilon_{norm}}{ N_{Norm} \epsilon_{sig}  },
\label{rd}
\ee
where ${\cal B}^i_{\tau}$ represents the branching fraction of the $i^{th}$ decay channel in which $\tau$ has been reconstructed. 
$N_{sig(norm)}$ and $\epsilon_{sig(norm)}$ represent the signal(normalization) events and the reconstruction efficiencies respectively.

On the other hand, the branching fraction in $B\to \tau\nu_{\tau}$ as defined in experimental analysis
is given by
\be
Br(B\to \tau\nu_{\tau})^i = \frac{N_s}{2 \epsilon^i_{\tau} N_{B^+B^-}},
\label{tau}\ee
where $\epsilon^i_{\tau}$ represents efficiency including the branching fraction of the $i^{th}$ decay mode of $\tau$, 
which is determined by the ratio of the number of events surviving all the selection criteria including the $\tau$ 
decay branching fractions to the number of fully reconstructed $B^{\pm}$. The total branching fraction is defined 
as the average of all the $Br(B\to \tau\nu_{\tau})^i$.

Now, $\tau$ decay channels include both the hadronic and leptonic final states. Therefore, we can define observables 
like $R_{\tau}^i = R(D^{(*)})^i/ Br(B\to \tau\nu_{\tau})^i$, for individual $\tau$ channels. For each of these ratios, any (unknown) systematics due to the $\tau$ identification is expected to cancel in the ratio $\epsilon^i_{\tau} / {\cal B}^i_{\tau}$. Then we can 
combine all such ratios to obtain the observable $R_{\tau}(D^{(*)})$.

\begin{figure*}[hbt]
\centering
 \includegraphics[scale=0.45]{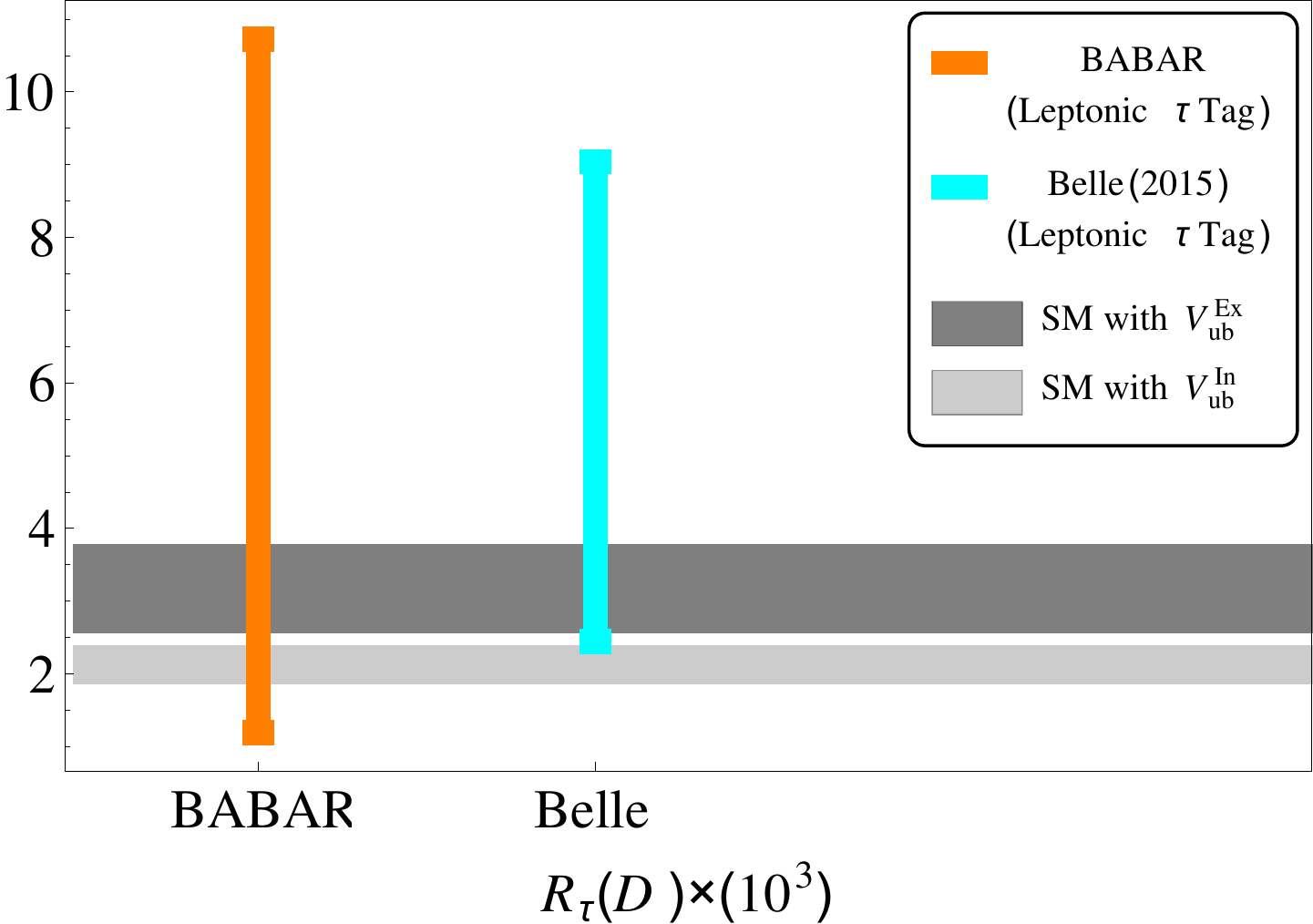}~~\includegraphics[scale=0.45]{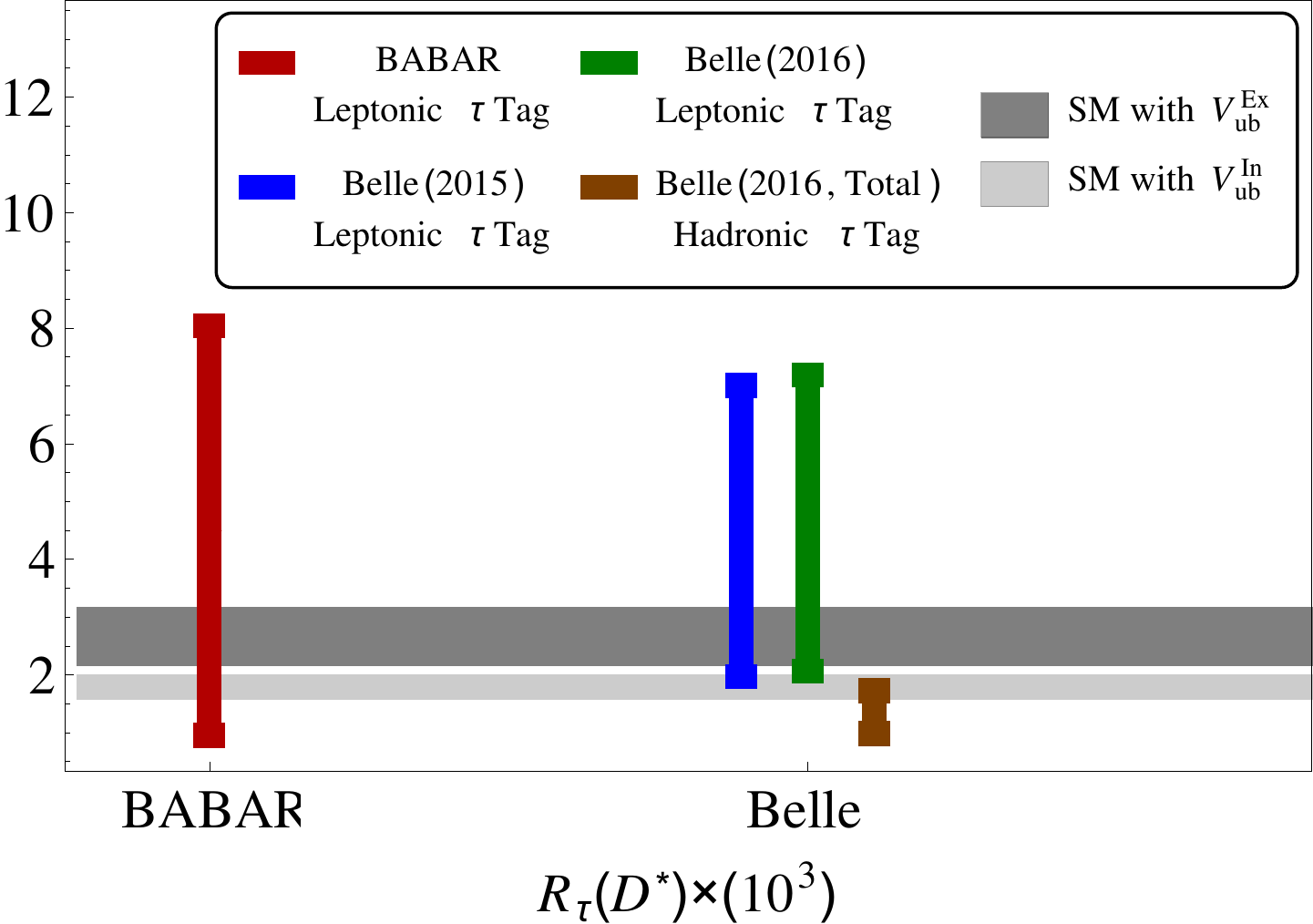}
 \caption{Graphical representation of the data shown in table \ref{tab:RtauDst}. Each experimental result of $\mathcal{R}(D^{(*)})$ is obtained either reconstructing the $\tau$s leptonically or hadronically. For calculation of $R_{\tau}(D^{(*)})$, along with the average value of $Br(B\to\tau\nu)$, we use the leptonic or hadronic average from tables \ref{tab:tauBABAR} and \ref{tab:tauBelle} depending on the $\tau$ reconstruction method for the specific $\mathcal{R}(D^{(*)})$.}
 \label{rnrnst}
\end{figure*}

As the available experimental results on $\mathcal{R}(D^*)$ do not mention the statistics for each 
channel separately, we show the values of $R_{\tau}(D^{(*)})$ in table \ref{tab:RtauDst} and Fig. \ref{rnrnst} using 
the average value of $Br(B\to\tau\nu)$ for either the leptonic or the hadronic channels depending on the $\tau$ detection procedure 
of the corresponding $\mathcal{R}(D^{(*)})$ measurement. For the rest of the paper, we will use use these leptonic or hadronic 
averages for $R_{\tau}(D^{(*)})$ unless otherwise specified.

While this observable (eq. \ref{rtaudeq}) has the advantage mentioned before of cancelling (unknown) $\tau$ detection systematics,
it has the drawback that it depends on $V_{ub}$. Consequently, this tends to increase the theory error in 
this observable but perhaps for testing the validity of the SM, this cautious approach has an advantage.   
Note also as stated earlier we tend to think that the exclusive $V_{ub}$ is now quite robust.

Be that as it may, the estimated values for $R_{\tau}(D^{(*)})$, using the results obtained in different experiments 
and the SM expectation, are listed in table \ref{tab:RtauDst} and plotted in Fig. \ref{rnrnst}. 
We note that all the estimated data from different experiments are consistent with the SM predictions;
remarkably, the 2 - 2.7$\sigma$ deviation of the \Babar~ results on 
$\mathcal{R}(D)$ and $\mathcal{R}(D^*)$ from the SM do not show up in our observable,
using inclusive or exclusive determinations of $V_{ub}$.
The \Babar~and Belle results are also fully consistent with each other, primarily due to the large errors in the data(particularly in the channels with leptonically tagged $\tau$s). In addition, $R_{\tau}(D^{(*)})$, obtained from channels with hadronically tagged $\tau$s has smaller error and differs from the SM prediction(with $V_{ub}^{Ex}$) by about $2.1 \sigma$. 

As the current measurements of $B \to \tau \nu$ have rather large errors, it may well be that for now the ratio is 
hiding NP beneath the errors; however, we are stressing its long term use as more data becomes available. 
Also, as alluded to before, we want to emphasize again that the consistency with the SM in Fig. \ref{rnrnst} or table \ref{tab:RtauDst}
does not necessarily mean that it rules out presence of new physics.
It just means that the effects of new physics, if there, largely cancel in the ratios. 
Later we will illustrate this with a particular example of new physics, {\it i.e} type-II 2HDM.

\subsection{$\mathcal{R}(X_c)$ and $R_{\tau}(X_c)$}\label{sec:RXcSM}

We now consider the inclusive decay channel $B\to X_c\tau\nu_{\tau}$ along with the exclusive channels
$\bdtau$, discussed earlier. If there is NP in $b\to c\tau\nu_{\tau}$, it should show up in  
both the exclusive and inclusive channels. Inclusive semileptonic decays are theoretically clean compared to the 
respective exclusive decays, but experimentally challenging. The forthcoming experiments like Belle-II may 
allow a precise measurement of the branching fraction of $B\to X_c\tau\nu_{\tau}$.
One potential advantage of the inclusive mode is that its branching ratio is
expected to be larger than the exclusive modes.  In a B-factory environment,
as in Belle-II, the experimental detection may be facilitated by (partial)
reconstruction of the ``other B". Unfortunately, at LHCb the inclusive
measurements are always very challenging.

The SM expression for the differential decay rate of inclusive $\overline{B} \to \tau \bar{\nu} X_c$ transitions 
including the power corrections at order $1/m_b^2$ in heavy quark effective theory (HQET) is\cite{Falk:1994gw,Ligeti:2014kia}

\begin{widetext}
\begin{align}\label{dgdq2Xc}
 \nn \frac{d \Gamma}{d \hat{q}^2} &= \frac{\left|V_{cb}\right|^2 G^2_F m^5_b}{192 \pi^3} \times 2 \left(1 - x_{\tau}\right)^2 \sqrt{P^2 - 4 \rho} \left\{\left(1 + \frac{\lambda_1 + 15 \lambda_2}{2 m^2_b}\right) \left[3 \hat{q}^2 P \left(1 + x_{\tau}\right) + \left(P^2 - 4 \rho\right) \left(1 + 2 x_{\tau}\right) \right] \right.\\
 & \left. + \frac{6 \lambda_2}{m^2_b} \left[(P-2)\left(1 + 2 x_{\tau}\right) + \hat{q}^2 
 \left(4 + 5 x_{\tau}\right) + \hat{q}^2 \frac{2 \left(2 \hat{q}^2 + P - 2\right) \left(2 + x_{\tau}\right) + 3 \hat{q}^2 P
 \left(1 + x_{\tau}\right)}{P^2 - 4 \rho} \right] \right\}
\end{align}
\end{widetext}

where, $\hat{q}^2 = q^2/m^2_b$, $q = p_{\tau} + p_{\nu}$  is the dilepton momentum, 
$x_{\tau} = m^2_{\tau}/ q^2 = \rho_{\tau}/\hat{q}^2$, $\rho = m^2_c / m^2_b$, $P = 1 - \hat{q}^2 + \rho$, $\lambda_1$ 
and $\lambda_2$ parametrize the leading non-perturbative corrections of relative order $1/m^2_b$. 
Integrating this over the range $\rho_{\tau} < \hat{q}^2 < (1 - \sqrt{\rho})^2$ gives us the total decay rate 
$\Gamma_W$.

Like $\mathcal{R}(D)$, the ratio of inclusive decay rates is defined as
\begin{equation}
\mathcal{R}(X_c) = \frac{\mathcal{B}(B \to X_c \tau \bar{\nu})}{\mathcal{B}(B \to X_c e \bar{\nu})}
\end{equation}
and its SM value, considering the current world average $\mathcal{B}(B^- \to X_c e \bar{\nu}) = (10.92 \pm 0.16)\%$   
\cite{Bernlochner:2012bc,Amhis:2012bh}, is given by $\mathcal{R}(X_c)_{SM} = 0.225 \pm 0.006$. 
In order to estimate $\mathcal{R}(X_c)_{Exp}$, we take the ratio of the LEP average   
$\mathcal{B}(b \to X_c \tau^+ \bar{\nu})_{LEP} = (2.41 \pm 0.23)\%$ \cite{Beringer:1900zz} and the 
world average for $\mathcal{B}(B^- \to X_c e \bar{\nu})$, and we obtain $\mathcal{R}(X_c)_{Exp} = 0.221 \pm 0.021$.

\begin{figure}[]
\centering
\includegraphics[scale=0.4]{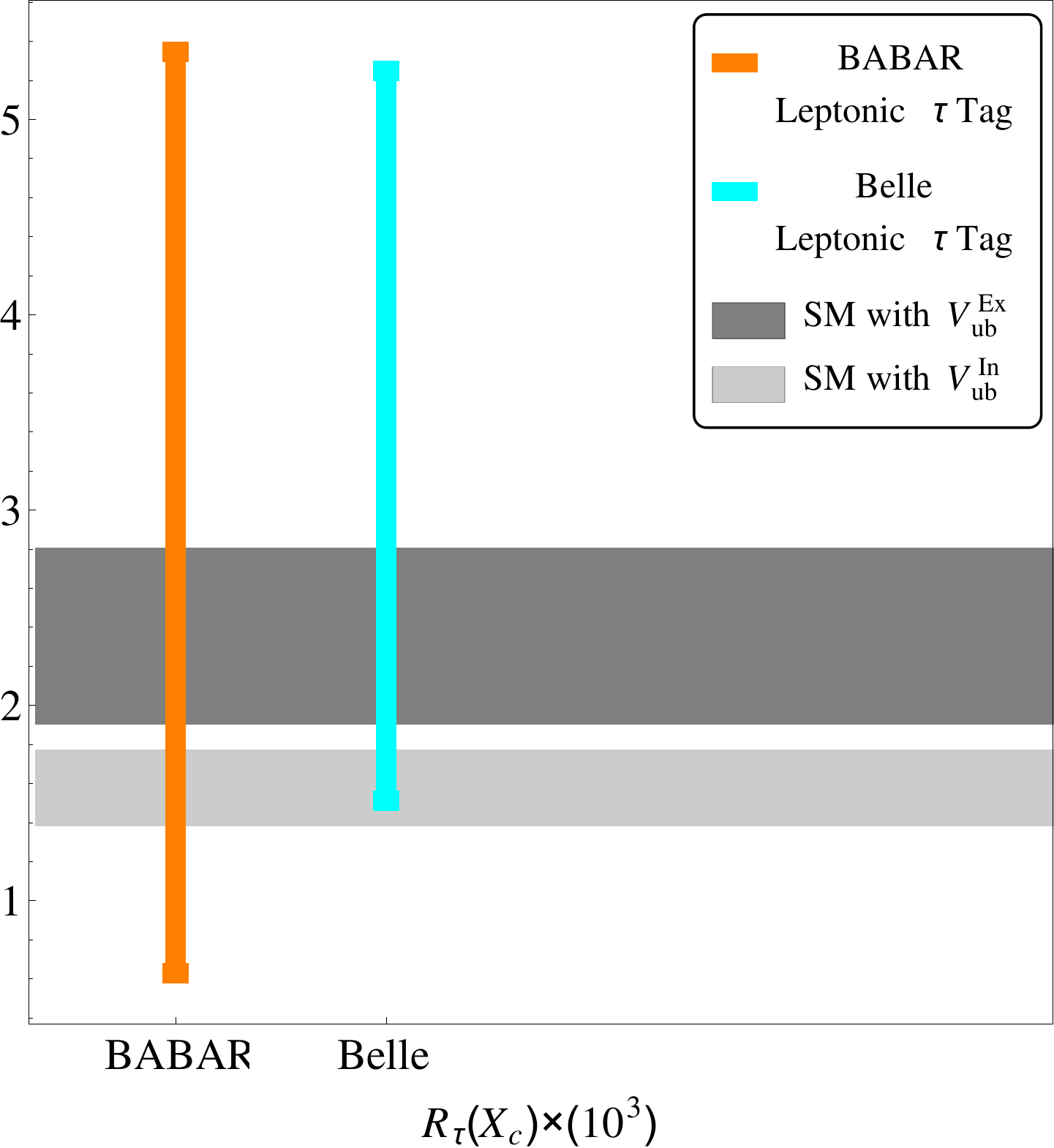}
\caption{SM (horizontal gray bands) estimation of $R_{\tau}(X_c)$ juxtaposed with experimental measurements (vertical bars).}
\label{RtXcCombo}
\end{figure}

Like $R_{\tau}(D^{(\ast)})$, we define $R_{\tau}(X_c)$ by normalizing $\mathcal{R}(X_c)$ with $\mathcal{B}(B^+ \rightarrow \tau^+ \nu_{\tau})$. In Fig. \ref{RtXcCombo}, different values of $R_{\tau}(X_c)$ are shown which are obtained for different values of the $\mathcal{B}(B^+ \rightarrow \tau^+ \nu_{\tau})$ taken from \Babar~and Belle measurements (for channels with  leptonically tagged $\tau$s 
in Fig. \ref{btn}). We note that the estimated values obtained using both the measurements are compatible with each other,  also both of them are consistent with the SM. Again, as in the case of exclusive modes this does not necessarily mean that
presence of all types of new physics are being ruled out.

\subsection{$\mathcal{R}(\pi)$ and $\mathcal{R}^{\pi}_{\tau}$ }\label{sec:Rpi}

Since $\mathcal{R}(D^{(*)})$ is independent of $|V_{cb}|$, if the interpretation of new physics there is correct, then we should expect similar deviations in analogous semileptonic decays$(B \to \pi (\rho,\omega)~\ell(\tau)~\nu)$ and similarly in $B_S$ decays. Therefore, in addition to the above modes we also consider the decay $B\to\pi\tau\nu_{\tau}$. Earlier, in the literature this mode is
considered for NP searches \cite{Tanaka:1994ay,Chen:2006nua,Kim:2007uq,Khodjamirian:2011ub,Bernlochner:2015mya}, 
while $B\to\pi\ell \nu_{\ell}$, with $\ell = \mu$ or $e$, is used for the extraction of CKM element $V_{ub}$ 
\cite{Bellevub,Babarvub}. The useful observable which is potentially sensitive to NP is defined as 

\be
\mathcal{R(\pi)} = \frac{\mathcal{B}(B \to \pi \tau \bar{\nu}_{\tau})}{\mathcal{B}(B \to \pi \ell \bar{\nu}_{\ell})},
\ee
where the dependence and therefore the uncertainty due to $V_{ub}$ cancels in the ratio; similarly, $\mathcal{R(\rho,\omega)}$ should also be studied. 

In the SM, the differential decay rate for the decay $B\to\pi\tau\nu_{\tau}$ is given as \cite{Bernlochner:2015mya}:
\begin{align}\label{dgamb2ptn}
 \nn \frac{d\Gamma(B \to \pi \tau \bar{\nu}_{\tau})}{d q^2} &= \frac{8 |\vec{p}_{\pi}|}{3} 
 \frac{G^2_F |V_{u b}|^2 q^2}{256 \pi^3 m^2_B} \left(1 - \frac{m^2_{\tau}}{q^2}\right)^2 \\
 & \left[H^2_0(q^2) \left(1 + \frac{m^2_{\tau}}{2 q^2}\right) + \frac{3 m^2_{\tau}}{2 q^2} H^2_t(q^2) \right]\,,
\end{align}
where, $q$ is the four-momentum transfer between the B-meson and the final-state pion of the semileptonic decay, 
$|\vec{p}_{\pi}|$ is the absolute three-momentum of the final state pion,
\begin{align}
 |\vec{p}_{\pi}| =\sqrt{\left(\frac{m^2_B + m^2_{\pi} -q^2}{2 m_B}\right)^2 - m^2_{\pi}}
\end{align}

and $H_{0/t}$ are helicity amplitudes defined as
\begin{align}
 \nn H_0 &= \frac{2 m_B |\vec{p}_{\pi}|}{\sqrt{q^2}} f_+(q^2)\\
  H_t &= \frac{m^2_B - m^2_{\pi}}{\sqrt{q^2}} f_0(q^2)\,.
\end{align}

The form factors $f_{+/0}$ need to be calculated using non-perturbative methods,
such as the lattice \cite{Flynn:2015mha,Lattice:2015tia}. 
Setting $m_{\tau}$ to zero in eq.(\ref{dgamb2ptn}) gives us the expression for 
$d\Gamma(B \to \pi \ell \bar{\nu}_{\ell}) / d q^2$ to an excellent precision. Taking the BCL coefficients and their correlations
from ref.\cite{Bernlochner:2015mya}, we calculate $R(\pi)^{SM} = 0.598 \pm 0.024$. The error is around 
4\%, which is only slightly larger than the value quoted in that paper ($0.641 \pm 0.016$) or essentially the same result of \cite{Du:2015tda}. 
Recent result from Belle \cite{Hamer:2015jsa} gives us an upper limit on $\mathcal{B}(B^0 \to \pi^- \tau^+ \nu_{\tau}) < 2.5 \times 10^{-4}$. 
Dividing this with the present world average of $\mathcal{B}(B^0 \to \pi^- \ell^+ \nu_{\ell}) = (1.45 \pm 0.05) 
\times 10^{-4}$ \cite{Agashe:2014kda}, we get the upper limit of $R(\pi) < 1.784$. 

We now introduce a different observable than our previous normalized ratio in eq.(\ref{rtaudeq})

\be\label{rtaupi}
 \mathcal{R}^{\pi}_{\tau} = \frac{\mathcal{B}(B \to \pi \tau \bar{\nu}_{\tau})}{\mathcal{B}(B \to \tau \nu_{\tau})}\,,
\ee
for which the SM prediction is $\mathcal{R}^{\pi,SM}_{\tau} = 0.733 \pm 0.144$; the error is around 20\%.  
We define $\mathcal{R}^{\pi}_{\tau}$ in this way instead of $\mathcal{R}(\pi) / \mathcal{B}(B^+ \rightarrow \tau^+ \nu_{\tau})$ 
as in the former definition the dependence due to $V_{ub}$ cancels. In the latter definition the dependence on 
$V_{ub}$ will remain, though the error in the SM is still around 20\%. Let us note in passing that a ratio analogous to eq.(\ref{rtaupi}) 
in case of $B \to D^{(*)} \tau \nu$ decays can also be useful.

Using the combined Belle result for $\mathcal{B}(B \to \tau \nu_{\tau})$(table \ref{tab:tauBelle}), and the upper 
limit for $R(\pi)$ quoted above, we obtain the upper limit for $\mathcal{R}^{\pi}_{\tau} < 2.62$. 

\section{Type II 2HDM Model}
The 2HDMs with two complex Higgs doublets are amongst the simplest extensions of the SM which gives rich phenomenology 
due to the additional scalar bosons. The extended Higgs sectors have not yet been ruled out experimentally.
The new features of the 2HDM includes three neutral and two charged Higgs bosons.    
The most general Yukawa Lagrangian induces flavour changing neutral current (FCNC) at the tree level. 
However, the 2HDM-II is designed to avoid FCNC at tree level. In this type, one Higgs doublet couples solely to 
up-type and the other one to down-type fermions \cite{Glashow:1976nt,Paschos:1976ay}. As a result the decay modes 
$B \to \tau \nu_{\tau}$ and $\bdtau$ are found to be sensitive to the effect of charged Higgs at the tree level.

In the following subsections we will discuss the constraints on 2HDM parameter space using the above mentioned observables.
The same analysis can be extended to other models as well. As we can see from Fig. \ref{fig_hfag}, $R(D)$ and $R(D^*)$ are 
highly correlated. Therefore, while constraining NP, it would not be a good idea to consider the data on
 $R(D)$ and $R(D^*)$ obtained using two different tagging methods. From now on, we will use the data given in table \ref{tab:RDst} and 
 \ref{tab:RtauDst}, except Belle's most recent(2016) result of $R(D^*)$ \cite{Abdesselam:2016cgx}, since, as explained before, in this measurement they have not yet given results on $R(D)$ using the same hadronic tag for ``the other B''.  


\begin{figure*}[]
\centering
 \includegraphics[scale=0.4]{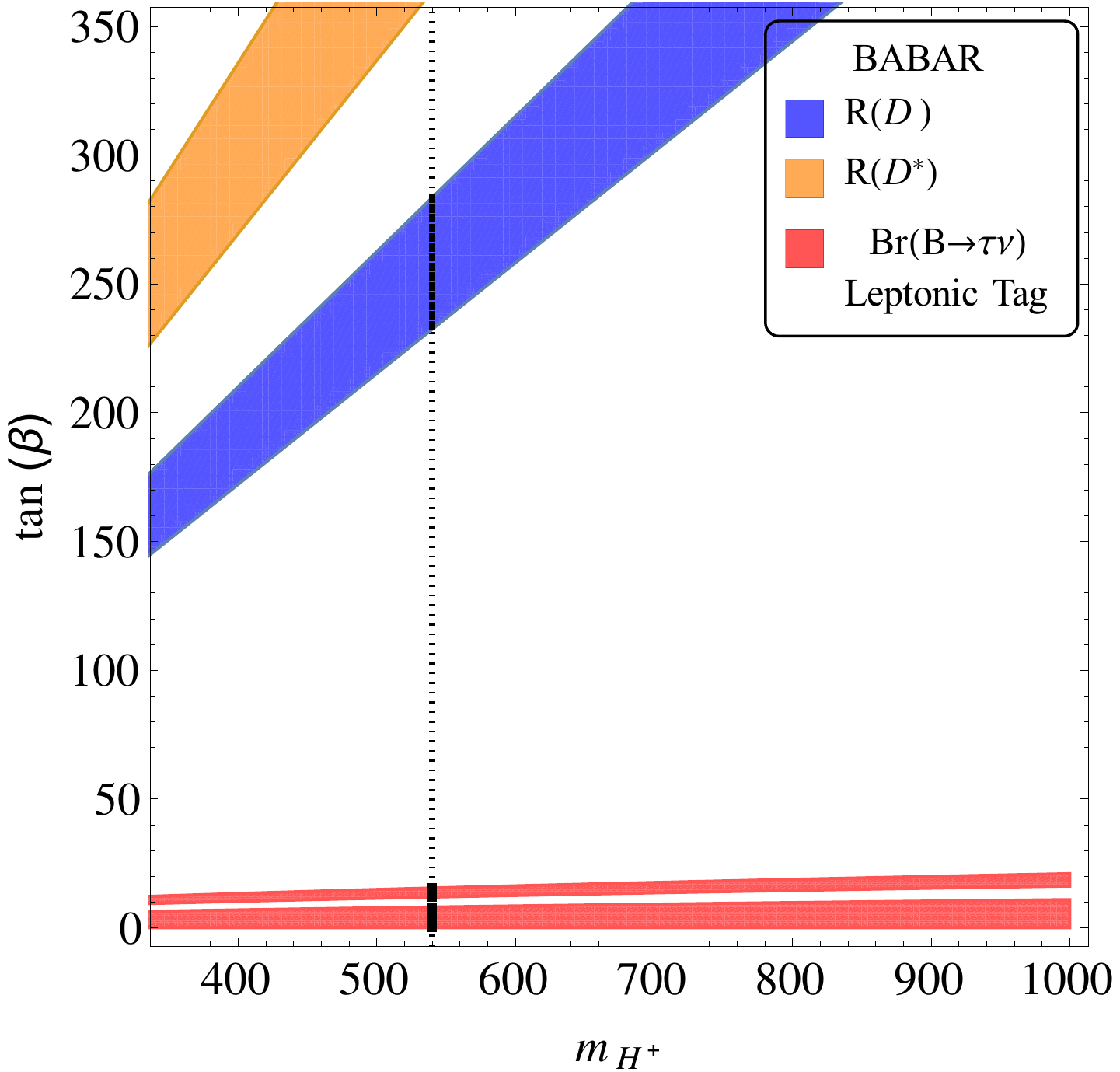}~~\includegraphics[scale=0.4]{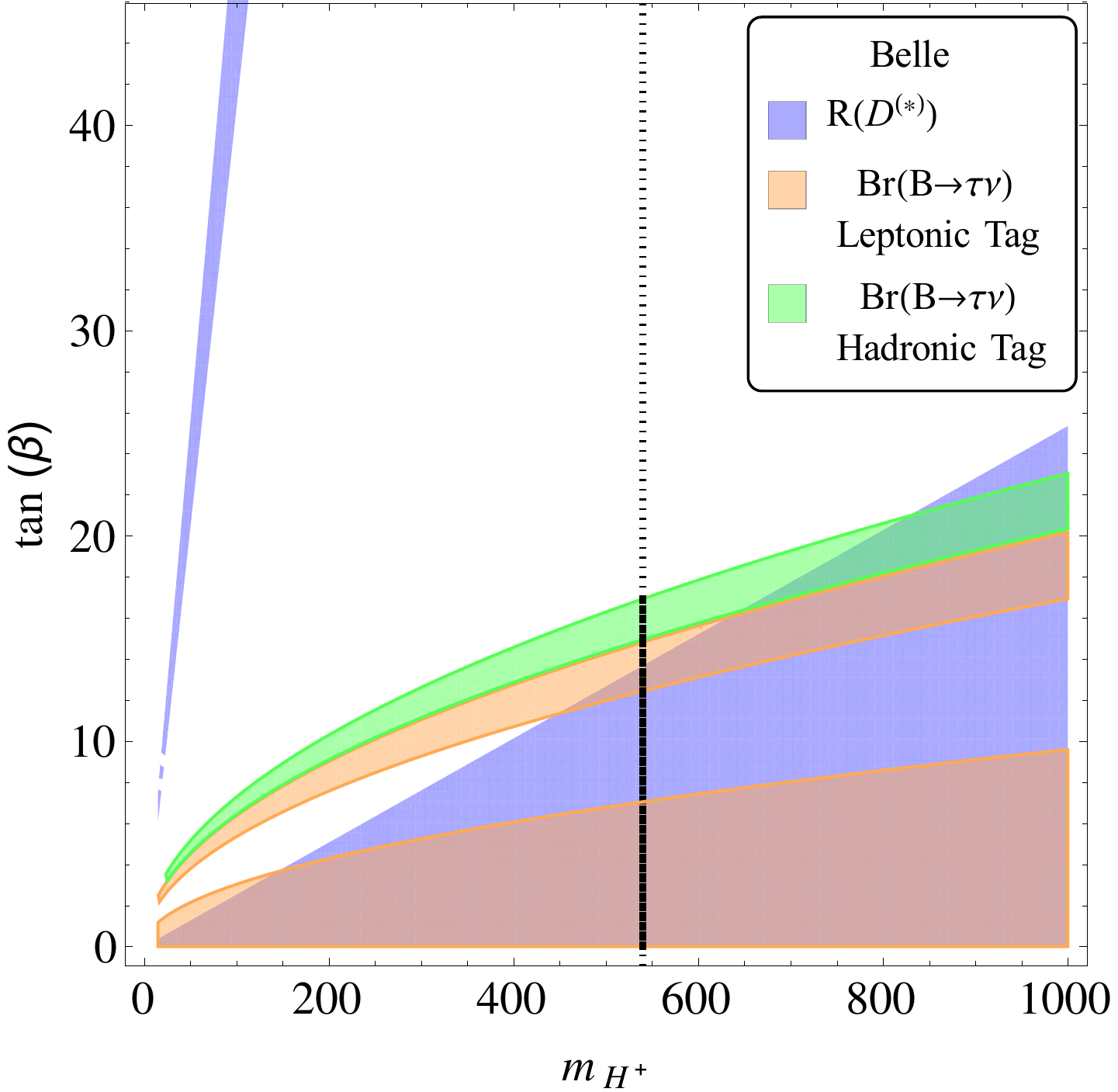}\\
 \caption{The allowed parameter spaces in 2HDM-II, which is obtained from $\mathcal{R}(D)$, $\mathcal{R}(D^*)$ and 
 $\mathcal{B}(B \to \tau \nu)$ using the \Babar~(left) and Belle (right) data. There is no common parameter space in 
 2HDM-II which satisfy simultaneously all the three excesses given by \Babar~. However, there are common parameter 
 spaces which are obtained as simultaneous solutions to all the three excess given by Belle data. The dotted vertical 
 lines show $m_{H^+} = 540$ GeV.}
 \label{RDbtnparamrange2HDm}
\end{figure*}

\begin{figure*}[]
\centering
 \includegraphics[scale=0.5]{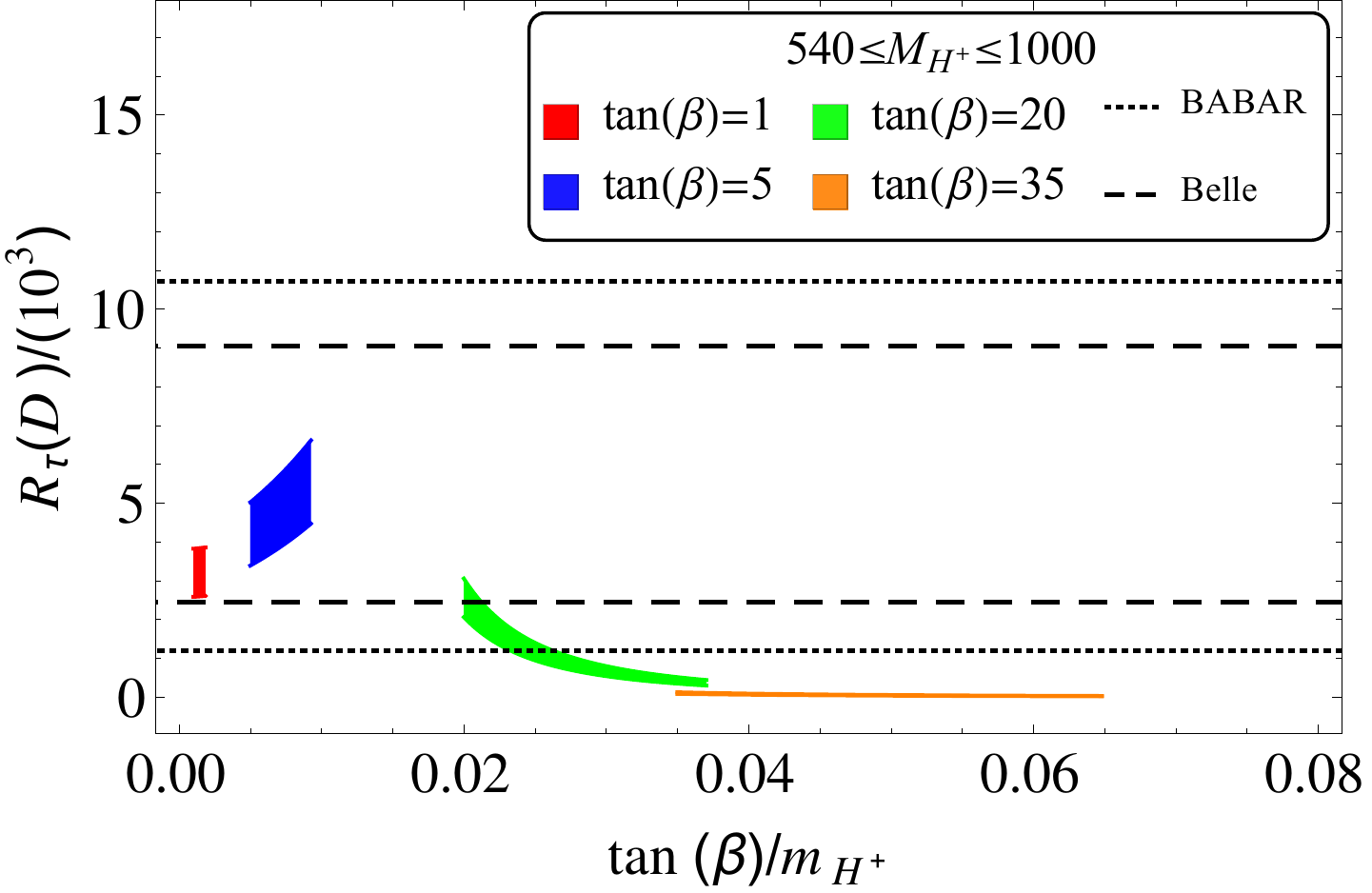}~~\includegraphics[scale=0.5]{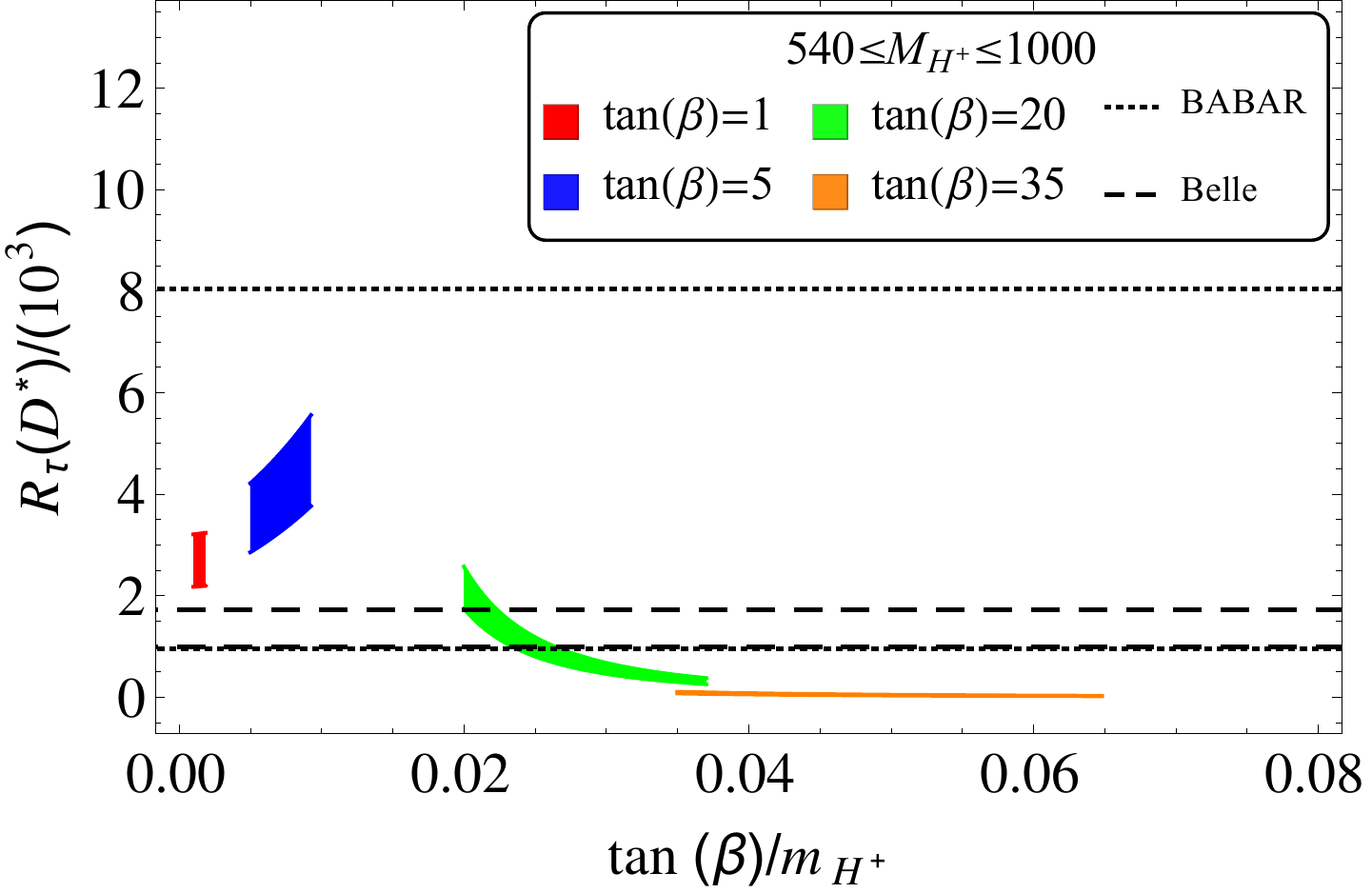}\\
 \caption{Variations of $R_{\tau}(D)$ (left) and $R_{\tau}(D^*)$ (right) with the 2HDM-II parameter 
 $r = \tan\beta/m_{H^+}$ for different values of $\tan\beta$. The $1\sigma$ experimental ranges are shown by 
 the dotted (\Babar~) and dashed (Belle) horizontal lines. }
 \label{rdtau2hdm}
\end{figure*}

\begin{figure*}[]
\centering
 \includegraphics[scale=0.4]{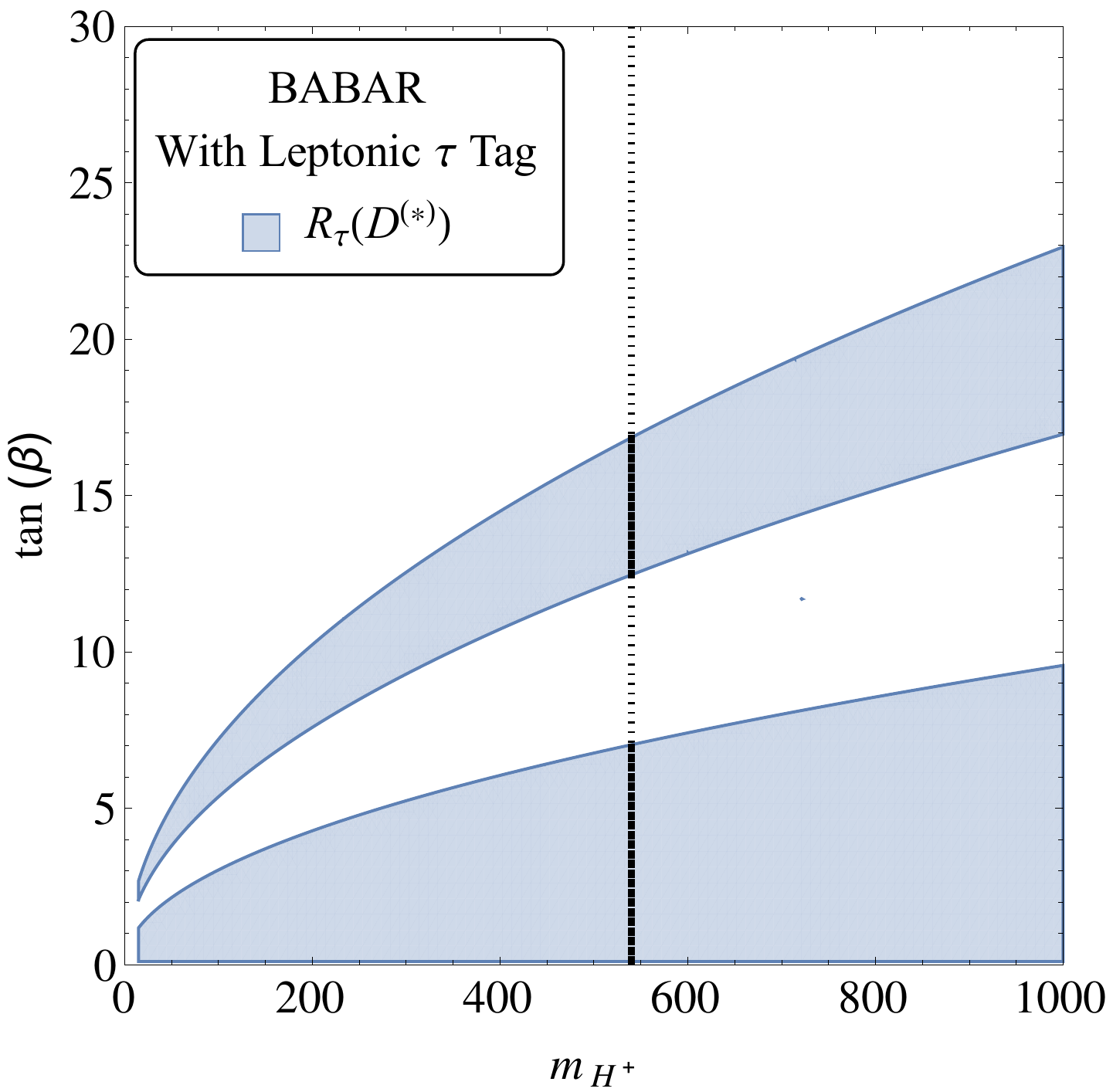}~~\includegraphics[scale=0.4]{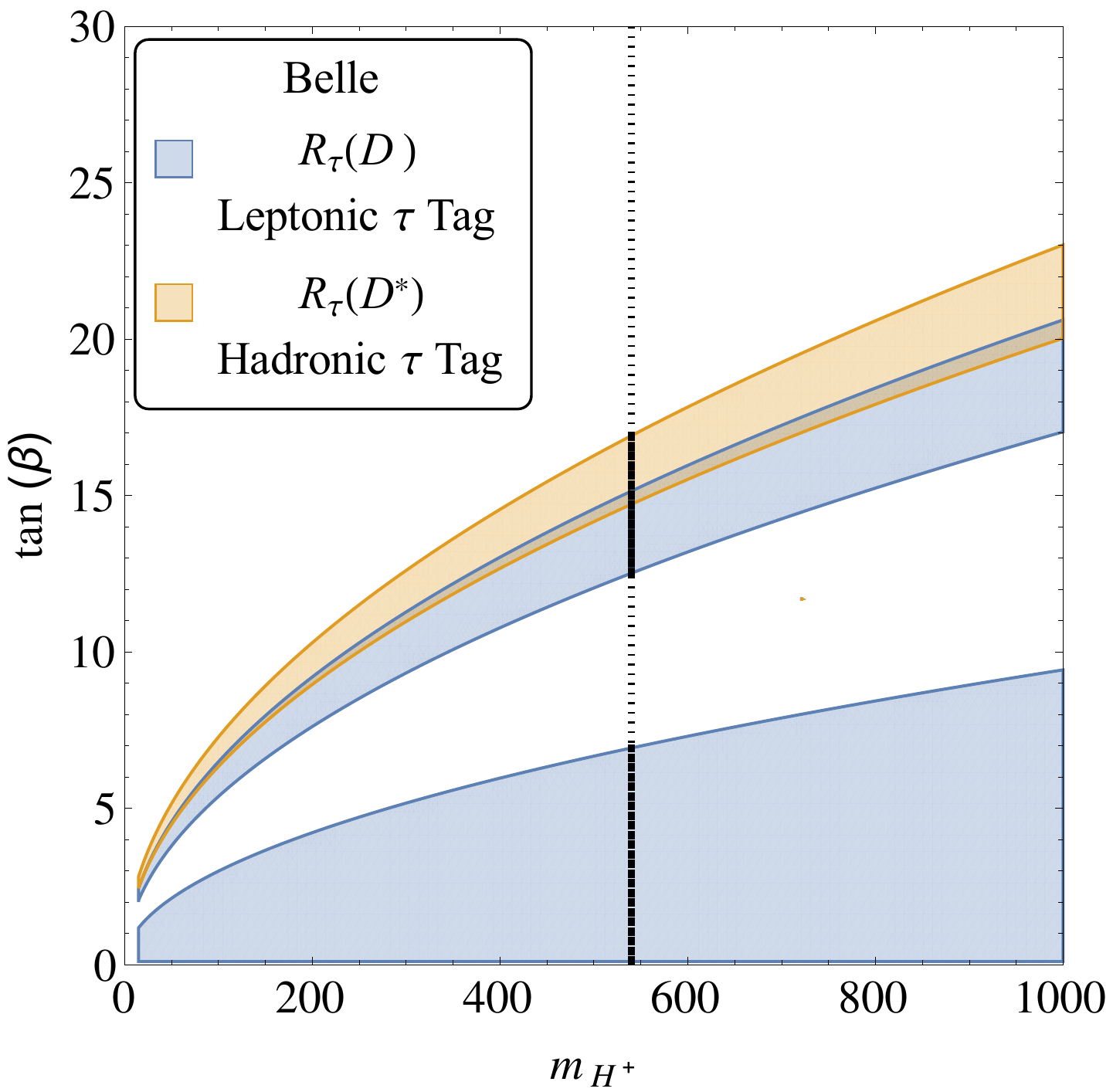}
 \caption{The allowed regions in $\tan\beta-m_{H^+}$ parameter space, which are obtained as simultaneous solutions to
 $\mathcal{R}_{\tau}(D)$ and $\mathcal{R}_{\tau}(D^*)$ using the data given in table \ref{tab:RtauDst}. The dotted vertical line shows $m_{H^+} = 540$ GeV.}
 \label{Rtauparamrange2HDm}
\end{figure*}  

\subsection{$\mathcal{B}(B^+ \rightarrow \tau^+ \nu_{\tau})$ and $\mathcal{R}(D^{(\ast)})$}

In two Higgs doublet models, purely leptonic decays receive an additional contribution from charged Higgs, 
which can be factorized from the SM prediction \cite{Hou:1992sy,Akeroyd:2007eh,Akeroyd:2003zr,Akeroyd:2003jb,Akeroyd:2009tn,Crivellin:2012ye},
\be\label{btn2HDM}
\mathcal{B}(B \to l \nu) = \mathcal{B}(B \to l \nu)_{SM} ~(1 + r_H)^2.
\ee
In 2HDM-II, the factor $r_H$ is given as, 
\be
r_H =\left(\frac{\left(m_{u}/m_{b}\right) - \tan^2{\beta}}{1 + \left(m_{u}/ m_{b}\right)}\right)
\left(\frac{m_B}{m_{H^+}}\right)^2 \,.
\ee
Here, $\tan{\beta}$ is the ratio of the vacuum expectation values of the two Higgs doublets, $m_{H^+}$ is 
the mass of the charged Higgs and $m_u/m_b = (0.56 \pm 0.06)\times 10^{-3}$\cite{Amsler:2008zzb} is the ratio of the $u$- and 
$b$-quark masses at a common mass scale.

The contributions of the charged Higgs to $\overline{B} \to D^{(*)} \tau^- \bar{\nu}_{\tau}$ decays can be encapsulated 
in the scalar helicity amplitude in the following way\cite{Kamenik:2008tj,Tanaka:1994ay}:

\be\label{Hs2HDM}
  H^{2HDM}_s \approx H^{SM}_s \times \left(1 - \frac{\tan^2{\beta}}{m^2_{H^{\pm}}} \frac{q^2}{1 \mp m_c / m_b}\right)\,.
\ee
The denominator of the second term of the above equation contain $(1 \mp m_c/m_b)$, where the negative and positive 
signs are applied to $\overline{B} \to D \tau^- \bar{\nu}_{\tau}$ and $\overline{B} \to D^{*} \tau^- \bar{\nu}_{\tau}$ 
decays, respectively. Here, $m_c / m_b = 0.215 \pm 0.027$ \cite{Xing:2007fb} is the ratio of the $c$- and $b$-quark 
masses at a common mass scale. Thus, the differential decay rate, integrated over angles, becomes 
\cite{Lees:2013uzd,Kiers:1997zt,Sakaki:2014sea},
\begin{align}
  \nn \frac{d\Gamma_{\tau}}{d q^2} &= \frac{G^2_F \left|V_{cb}\right|^2 \left|{\bf p}^*_{D^{(*)}}\right| q^2}
  {96 \pi^3 m^2_B} \left(1 - \frac{m^2_{\tau}}{q^2}\right)^2 \left[\left(\left|H_+\right|^2 \right. \right.\\
  &\left.\left. + \left|H_-\right|^2 + \left|H_0\right|^2 \right) \left(1 + \frac{m^2_{\tau}}{2 q^2}\right) + 
  \frac{3 m^2_{\tau}}{2 q^2} \left|H_s\right|^2\right]
\end{align}
where $\left|{\bf p}^*_{D^{(*)}}\right|$ is the three-momentum of the $D^{(*)}$ meson in the $B$ rest frame. 
Given that charged Higgs bosons are not expected to contribute significantly to $\overline{B} \to D l^- \bar{\nu}_{l}$ 
decays, $\mathcal{R}\left(D^{(*)}\right)_{2HDM}$ can be described by a parabola,

\begin{align}\label{RD2hdm}
  \nn \mathcal{R}\left(D^{(*)}\right)_{2HDM} &= \mathcal{R}\left(D^{(*)}\right)_{SM} + A_{D^{(*)}} 
  \frac{\tan^2{\beta}}{m^2_{H_+}} \\
  &+ B_{D^{(*)}} \frac{\tan^4{\beta}}{m^4_{H_+}}\,,
\end{align}
where,
\begin{align}
  \nn A_{D} &=  -3.25 \pm 0.32, ~~ A_{D^{*}} = -0.230 \pm 0.029 \\
  \nn B_{D} & = 16.9 \pm 2.0, ~~~~ B_{D^{*}} = 0.643 \pm 0.085\,.
\end{align}

$A(B)_{D^{(*)}}$ are determined by averaging over $B^0$ and $B^−$ decays\cite{Lees:2013uzd}. The uncertainty 
estimation includes the uncertainties on the mass ratio $m_c /m_b$ and the form factors, as well as their correlations.

The allowed parameter space in the $\tan{\beta}$ - $m_{H^+}$ plane using 
the excess observed by \Babar~and Belle on $\mathcal{B}(B^+ \rightarrow \tau^+ \nu_{\tau})$ and $\mathcal{R}(D^{(\ast)})$, 
are shown in Fig. \ref{RDbtnparamrange2HDm}. We note that \Babar~data do not allow a simultaneous explanation of all 
the above mentioned deviations. However, for Belle data, there actually is a common allowed parameter space and we show 
that by overlapping the regions allowed by $\mathcal{R}(D^{(*)})$ and $\mathcal{B}\left(B \to \tau \nu\right)$ 
The regions for $m_{H^+} \le 540$ GeV have 
already been excluded by $b \to s \gamma$ measurements at 95\% confidence level \cite{Pesantez:2015qjc}. 
In Fig.\ref{RDbtnparamrange2HDm} and onwards, this bound is shown as a dotted line.

In the 2HDM-II model, the variations of $R_{\tau}(D^{(*)})$ with $r = \tan{\beta}/m_{H^+}$ for various values of 
$\tan\beta$, while $m_{H^+}$ is being kept in between $[540,1000]$, are shown in Fig. \ref{rdtau2hdm}. 
We note that both \Babar~and Belle data discard a solution with large $\tan\beta$ ($\gsim 30$). The allowed parameter space in  
$\tan{\beta}$ - $m_{H^+}$ plane using the experimental constraints are shown in Fig. \ref{Rtauparamrange2HDm}.
Also both \Babar~and Belle data, given in table \ref{tab:RtauDst}, 
allow non-zero values of $\tan\beta$ and $m_{H^+}$, while large values of $\tan\beta$
are not allowed by the data. In addition, the allowed parameter space in all the different 
cases discussed in table \ref{tab:RtauDst} are consistent with each other.

\subsection{$\mathcal{R}(X_c)$ and $R_{\tau}(X_c)$}

\begin{figure}[]
\centering
 \includegraphics[scale=0.4]{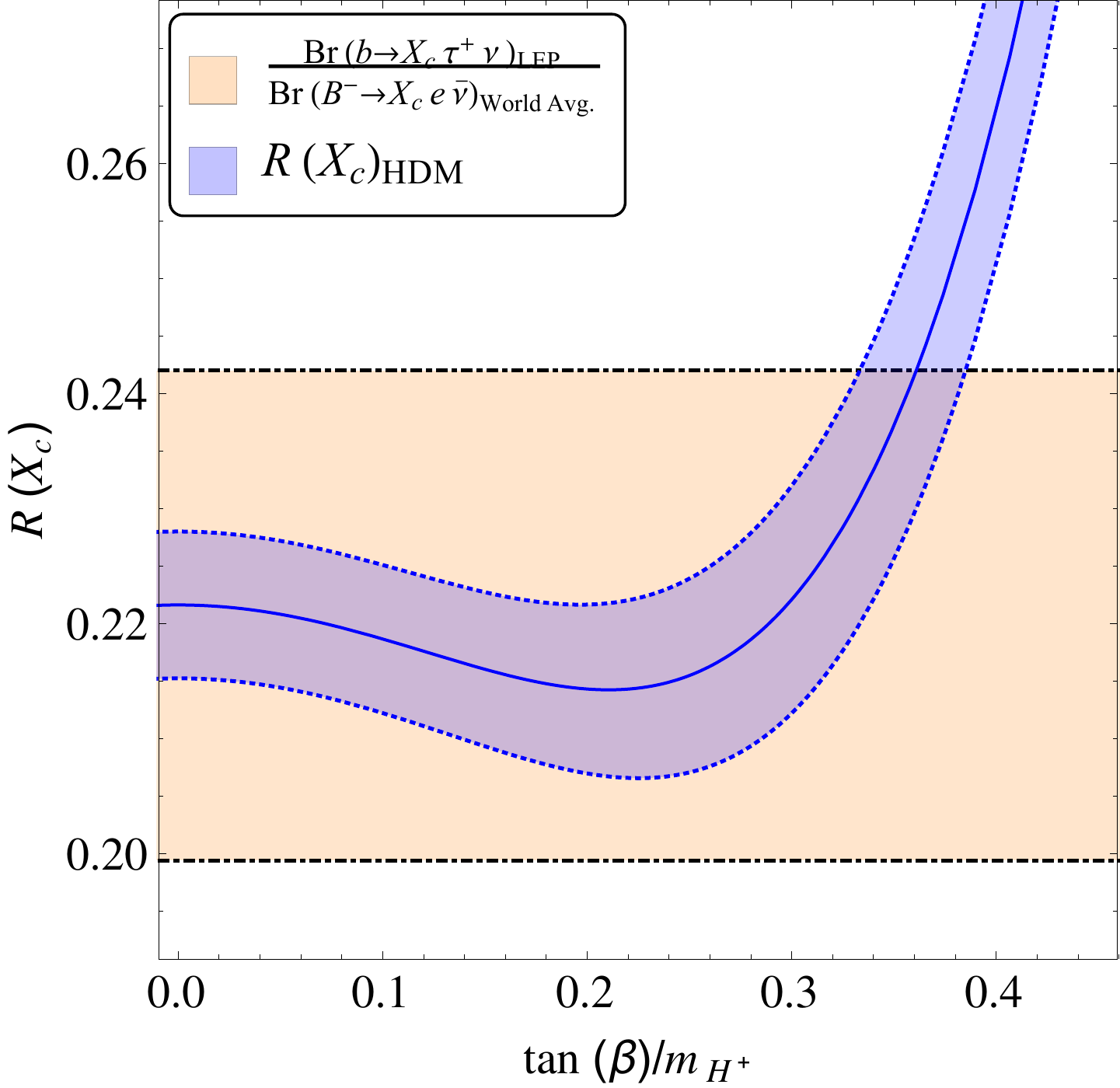}
 \caption{Variation of $R(X_c)$ with $r =\tan{\beta}/m_{H^+}$ (blue region between dotted curves). 
 Experimental range is shown by the orange region enclosed by dot-dashed horizontal lines. $r = 0$ corresponds to SM prediction for $R(X_c)$.}
 \label{RXcplt2HDM}
\end{figure}

The theoretical expression for total decay rate of inclusive $\overline{B} \to X_c \tau \bar{\nu}$ transitions including 
power corrections of order $1/m_b^2$ in HQET, in the framework of type II 2HDM model is given by \cite{Grossman:1994ax},

\be
\Gamma = \frac{\left|V_{cb}\right|^2 G^2_F m^5_b}{192 \pi^3}\left[\Gamma_W + \frac{R^2}{4} \Gamma_H - 
2 R \frac{m_{\tau}}{m_b} \Gamma_I\right]
\ee
where, $R = r^2 m_{\tau} m_b$, $r = \tan{\beta}/m_{H^+}$, and the subindices $W$, $H$, and $I$ denote the $W$ 
mediated(SM), Higgs mediated and interference contributions, respectively. $\Gamma_W$ is given by the $\hat{q}^2$ 
integrated form of eq.(\ref{dgdq2Xc}). Other terms are listed in ref. \cite{Grossman:1994ax}. 

\begin{figure}[]
\centering
  \includegraphics[scale=0.5]{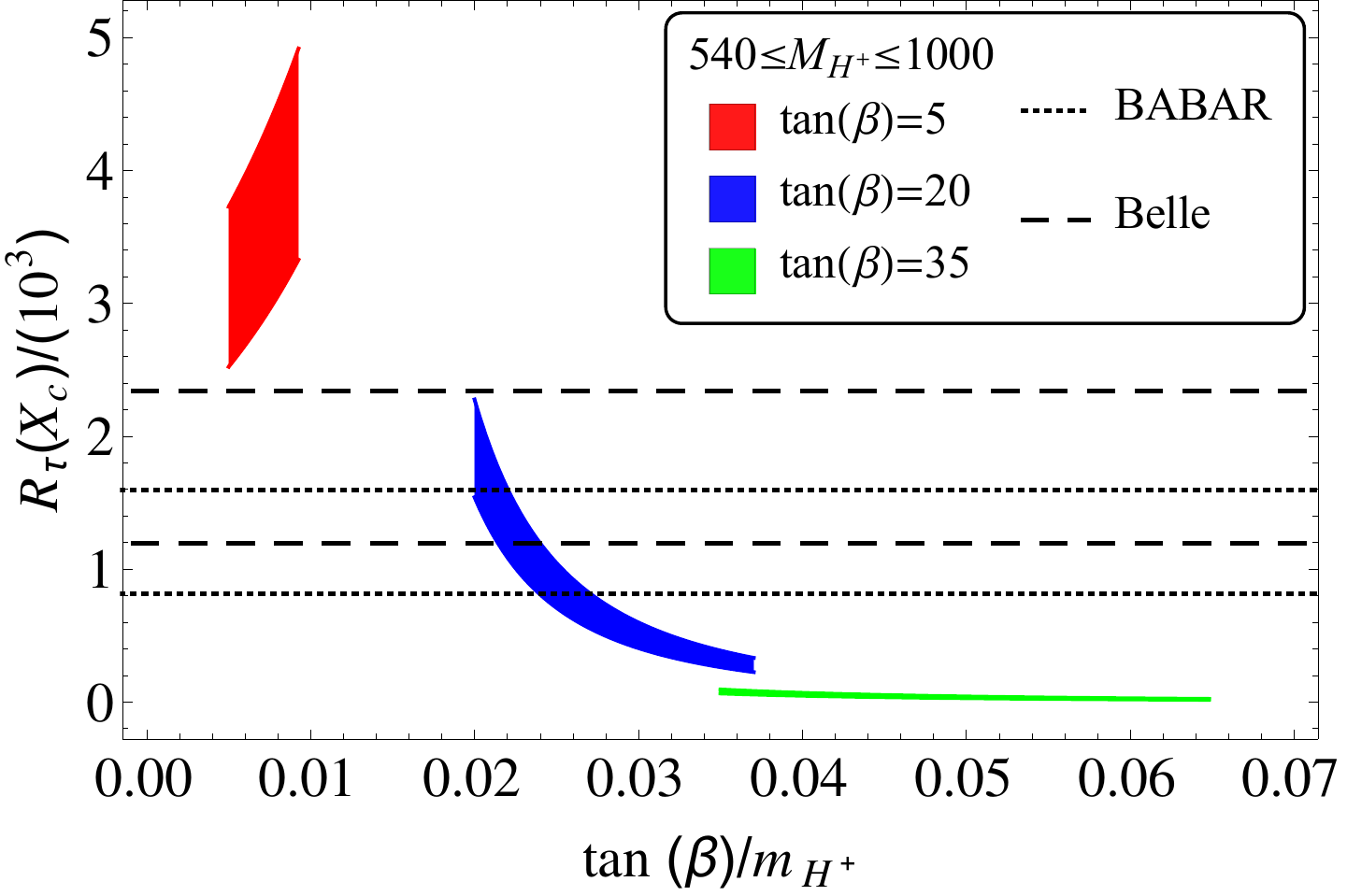}
  \caption{Variation of $R_{\tau}(X_c)$ with $r =\tan{\beta}/m_{H^+}$ for different values of $\tan\beta$ while 
  $380 <m_{H^+} < 1000$. Experimental ranges are shown by dotted (\Babar~) and dashed (Belle) horizontal lines .}
  \label{Rxctau2hdm}
 \end{figure}
Figure(\ref{RXcplt2HDM}) represents the variation of $\mathcal{R}(X_c)$ with $r$. We note that the current 
data allows only the region  $r \le 0.4$, the region $r >0.4$ is not allowed by the data.   

\begin{figure}[]
\centering
 \includegraphics[scale=0.4]{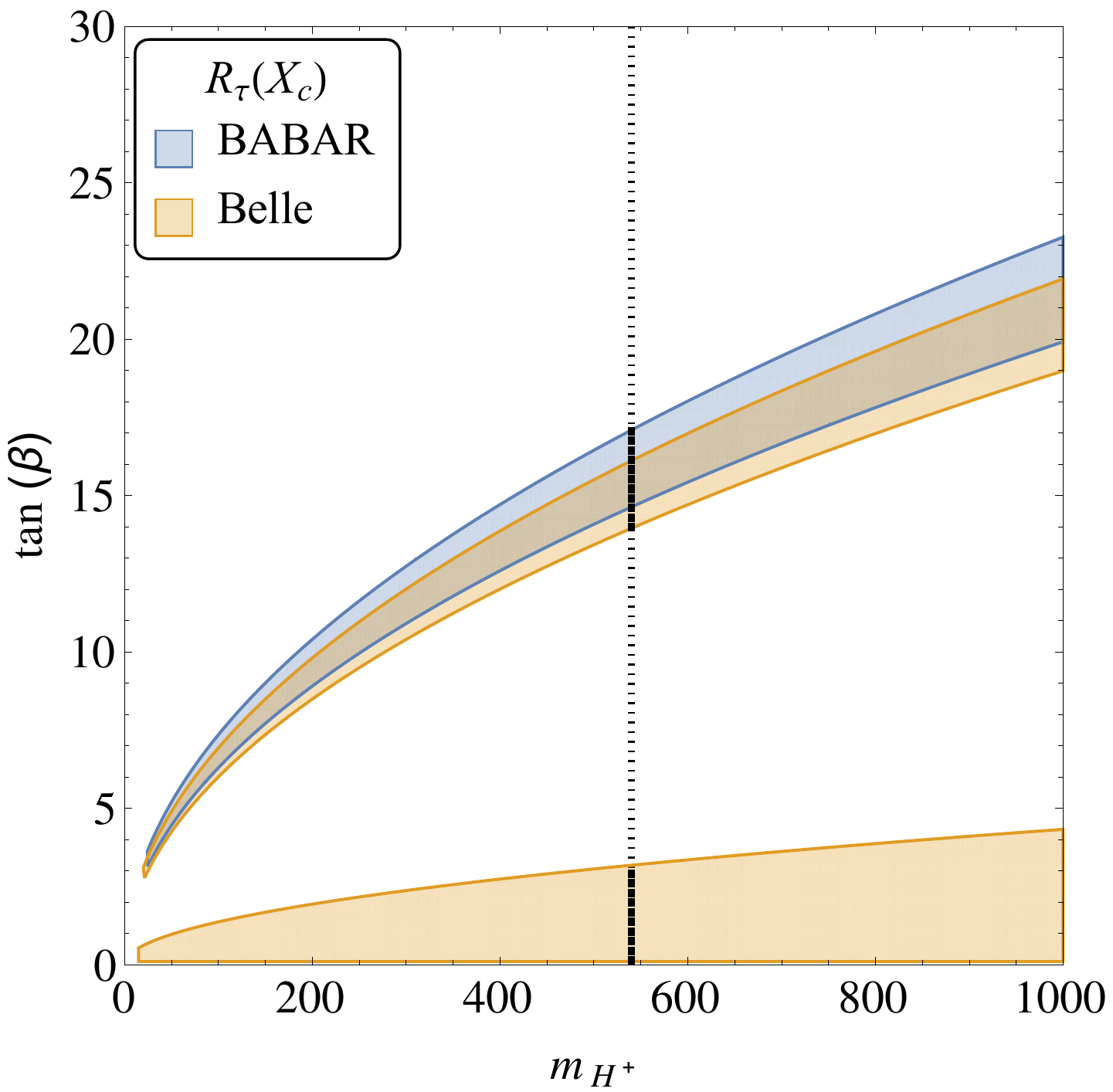}
 \caption{Allowed parameter space for $\tan\beta$ and $M_{H^+}$ obtained from the analysis of $R_{\tau}(X_c)$ in 2HDM-II. The dotted vertical line shows $m_{H^+} = 540$ GeV.}
 \label{RXctauplt}
\end{figure}

The variations of $\mathcal{R}_{\tau}(X_c)$ in 2HDM-II with the parameter $r$ for 
various values of $\tan\beta$ are shown in Fig. \ref{Rxctau2hdm}. Here too the $m_{H^+}$ is varied in 
between $[540,1000]$ as before. Also, in this case we note that the large values of $\tan\beta$ ($\gsim 30$) 
are not allowed by the current data. The experimental constraints in the $\tan\beta - m_{H^+}$ plane, obtained from the analysis of this observable is shown in Fig. \ref{RXctauplt}.   
 
\subsection{$\mathcal{R}^{\pi}_{\tau}$}

\begin{figure}[]
\centering
  \includegraphics[scale=0.5]{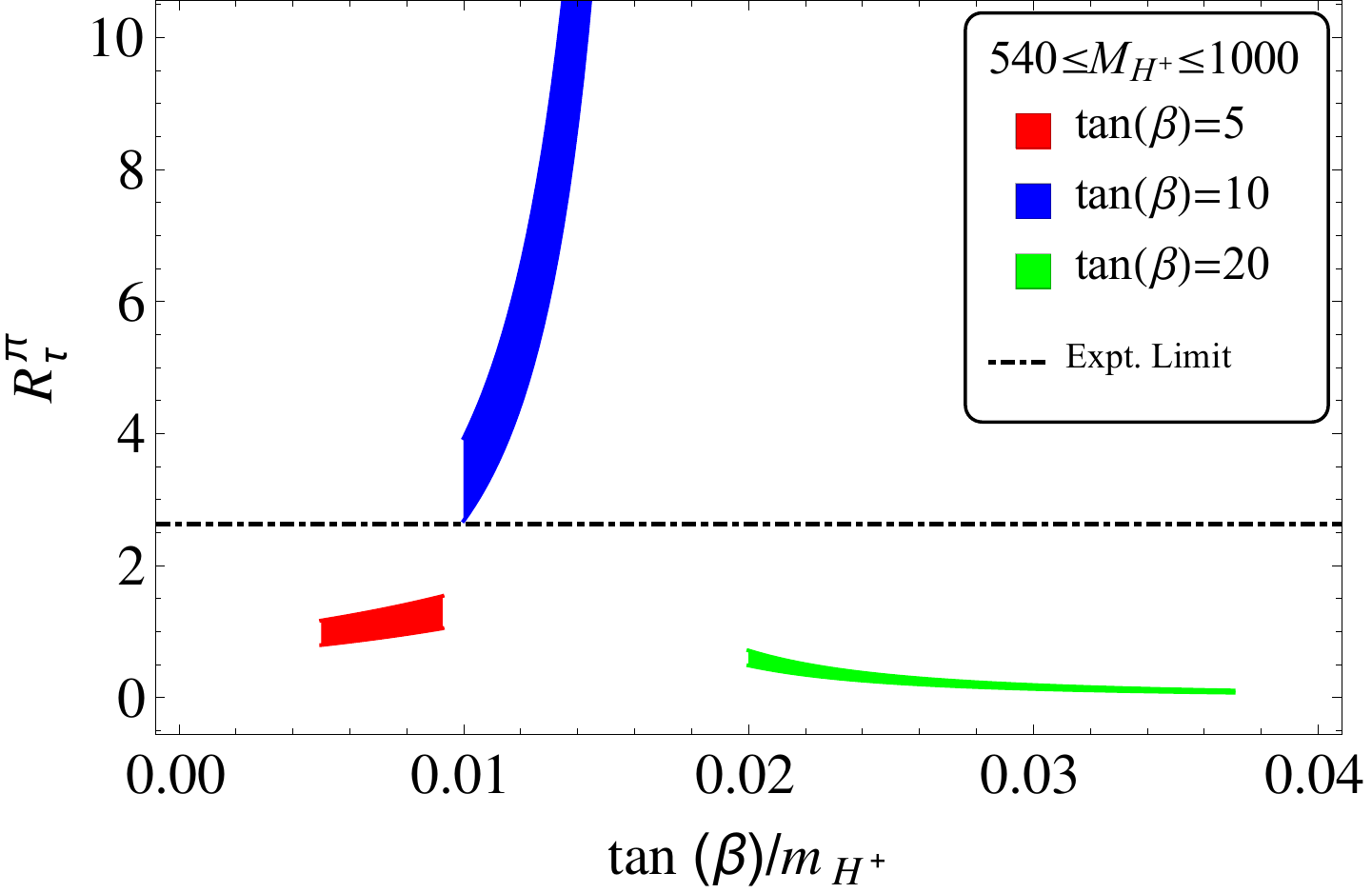}
  \caption{Variation of $\mathcal{R}^{\pi}_{\tau}$ with $r =\tan{\beta}/m_{H^+}$ for different values of $\tan\beta$ while 
  $540 <m_{H^+} < 1000$. Experimental upper limit(section \ref{sec:Rpi}) is shown by the dot-dashed horizontal line.}
  \label{Rpitau2hdm}
 \end{figure}

The contributions of the charged Higgs Boson to $B \to \pi \tau \bar{\nu}_{\tau}$ decays can be incorporated into 
Eq.(\ref{dgamb2ptn}) by the replacement\cite{Tanaka:1994ay,Kamenik:2008tj}
\be\label{Ht2HDM}
  H_t \to H^{SM}_t \times \left(1 - \frac{\tan^2{\beta}}{m^2_{H^{+}}} \frac{q^2}{1 - m_u / m_b}\right)\,.
\ee

So, just like eq.(\ref{RD2hdm}), $\mathcal{B}(B \to \pi \tau \bar{\nu}_{\tau})$ can be described as a parabola,
\begin{align}\label{b2ptn2hdm}
  \nn \mathcal{B}(\pi)_{2HDM} &= \mathcal{B}(\pi)_{SM} + A_{\pi} \frac{\tan^2{\beta}}{m^2_{H_+}} \\
  &+ B_{\pi} \frac{\tan^4{\beta}}{m^4_{H_+}}\,,
\end{align}
where,
\begin{align}
  \nn A_{\pi} &=  (-0.389 \pm 0.164) \times 10^{-3} \\
  B_{\pi} &= (0.418 \pm 0.258) \times 10^{-2}
\end{align}

\begin{figure}[]
\centering
 \includegraphics[scale=0.4]{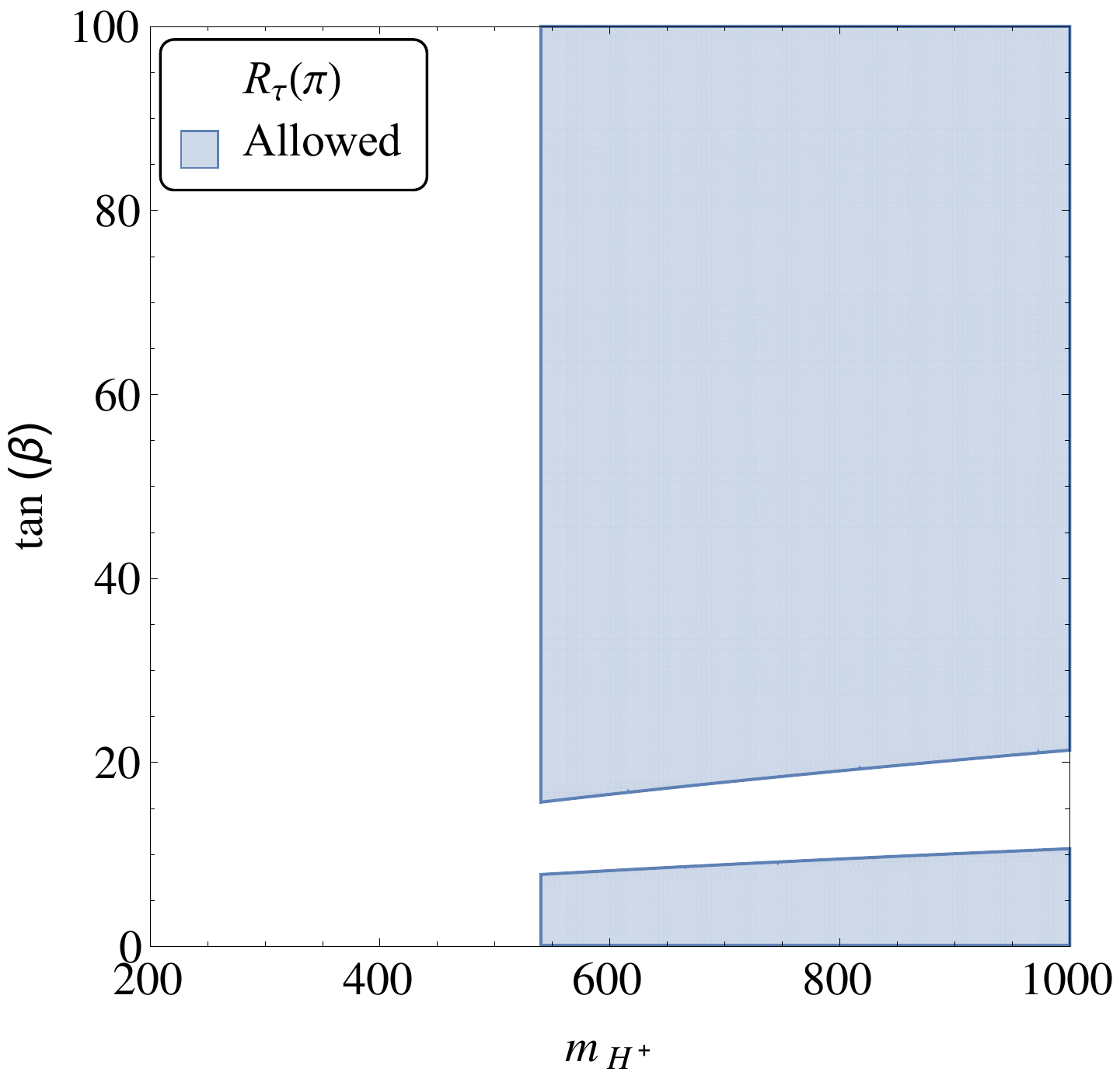}
 \caption{Allowed parameter space for $\tan\beta$ and $M_{H^+}$ obtained from the analysis of $R_{\tau}(\pi)$ in 2HDM-II.}
 \label{Rpitauplt}
\end{figure}

Figures (\ref{Rpitau2hdm}) and (\ref{Rpitauplt}) show the Type-II 2HDM parameter space corresponding to the 
experimental upper limit given in section(\ref{sec:Rpi}). We note that in this case $\tan\beta$ as large as 100 is allowed 
by the current data. 

\section{Summary \& Outlook}
Motivated by the reported indications of new physics signals in the
experimental results from \Babar, Belle and LHCb in the ratio, 
$\mathcal{R}(D^{(*)})$, of semileptonic decays,  in here we examine them along with $B\to \tau\nu_{\tau}$ decays.
Since $\tau$ detection plays a central role in both categories, 
and because backgrounds in $B \to D^{(*)} \tau \nu$ are very different from those in $B \to \tau \nu$, it seems very 
useful to examine them both simultaneously whenever the data allows. 
Concretely, we define a new observable, $R_{\tau}(D^{(*)}) \equiv \mathcal{R}(D^{(*)})/\mathcal{B}(B\to\tau\nu_\tau)$. 
In this observable the (unknown) systematics, if any, due to the $\tau$ identification are expected to largely cancel. 
Our analysis shows that this observable appears to be remarkably consistent with SM with the 
data from both \Babar~ and Belle \footnote{We cannot construct such a ratio for LHCb
as so far at LHCb it has been difficult to measure the branching ratio for
$B\to \tau\nu_{\tau}$.} even though appreciable differences from
the SM were reported in  $\mathcal{R}(D^{(*)})$ especially by \Babar. Since at present the errors in $B \to \tau \nu$ 
are rather large, it is certainly plausible that NP is for now hiding in the errors; our main purpose was to explore and suggest its use in the long run.

We emphasize that consistency of the experimental results on $\mathcal{R}_{\tau}(D^{(*)})$ with the SM does not 
necessarily mean the absence of all new
physics contribution in $\mathcal{R}(D^{(*)})$ and/or $B\to \tau\nu_{\tau}$. 
For example, for a class of new physics models which affect both type of decay modes (as happens in 2HDM-II), 
NP contributions would also largely tend to cancel in $\mathcal{R}_{\tau}(D^{(*)})$.  In fact our analysis 
of $\mathcal{R}_{\tau}(D^{(*)})$ explicitly  shows that type II-2HDM in the region of $M_H$ larger than about 
500 GeV with tan$\beta$ less than about 25 is allowed; in this important respect we reach at a different
conclusion than the \Babar~ analysis. Indeed, the constraint obtained on the parameter space in 
$\tan\beta$-$m_{H^+}$ plane is very similar to the one obtained from $\mathcal{B}(B\to\tau\nu_\tau)$; 
$\tan\beta \gsim 25$ is not allowed by the present data in the case when $m_{H^+} < 1$ {\it TeV}.

This conclusion regarding the possible relevance of type-II 2HDM is of 
special significance to Super-symmetric theories as therein type II-2HDM
are crucial.

Analogously, we also study inclusive semileptonic decays of final states
with $\tau$ in them. Experimentally,  these inclusive final states are especially challenging.
However, at least for Belle II by (partial) re-construction
of the ``other B", this may have a chance and thereby one may be able to
use the higher branching ratio of the inclusive mode. For this
purpose we define $R_{\tau}(X_c)$ for the inclusive decays $b\to X_c\tau\nu_{\tau}$, and find the constraints 
on 2HDM-II parameter space. Here too, the allowed regions are mostly dominated by the constraints from  
$\mathcal{B}(B\to\tau\nu_\tau)$. In addition to the above observable we also study the ratio 
$\mathcal{B}(B\to \pi\tau\nu_\tau)/\mathcal{B}(B\to\tau\nu_{\tau})$. At present, only 
an upper limit exists for this ratio, therefore we do not get strong bounds on $\tan\beta$ or $m_H^+$ plane from this observable. Also, large 
values of $\tan\beta \approx 100$ are still allowed. Here too, more precise measurements are necessary to probe the presence
for new physics. Also, there are  models of new physics which effect the two modes in the ratio differently. Our newly defined 
observable may play an important role in distinguishing the possible signature of those new physics  models from others. 

For a reliable interpretation of new physics signals in semi-tauonic final states 
in B-decays, not only experimental and theoretical precision in those modes is needed but also
more precise measurements of $\mathcal{B}(B\to\tau\nu_\tau)$ would help
significantly. The expected larger data samples at Belle II by factors of
around 25 to 50 should prove to be very useful in the next few years in this regard\footnote{Better determination of 
 $\mathcal{B}(B\to\tau\nu_\tau)$ would be very valuable for other reasons
as well~\cite{Lunghi:2009ke}.}.
Needless to say more accurate measurements of semi-tauonic modes would
also go a long way. Therein not only Belle-II but more data from LHCb should 
be forthcoming and should be very helpful in significantly reducing the
current uncertainties.  In all these interpretations lattice calculations
play a crucial role and in the next few years in many quantities of interest 
to all this physics,  percent or even sub-percent precision is, fortunately, anticipated. In this context, we want to reiterate that in lattice calculations, in the SM, in addition to $R(D^{(*)})$, ratios analogous to $R_{\tau}(D^{(*)})$ (see eq.(\ref{rtaudeq})), $R(\pi)$,  $R_{\tau}^{M^{(*)}}$ (see eq.(\ref{rtaupi})) for $B$, $B_S$ decays, with appropriate choices of $M^{(*)}$ relevant to charge-current semileptonic transitions, are desirable. This would ensure that correlations in lattice data are properly taken into account to enhance precision.


\section*{Acknowledgement}
This work is an outgrowth of Workshop of High Energy Physics Phenomenology (WHEPP XIV) in Kanpur, India (Dec 15); 
the authors thank the organizers for this opportunity. The work of AS is supported in part by the US DOE Contract
No. DE-SC 0012704.


\section*{References}

\begin{thebibliography}{99}

\bibitem{belle_talk}
 Y.~Sato [Nagoya Univ.],
 ``Tree-level New Physics searches in semileptonic B decays at Belle''
  \href{https://indico.cern.ch/event/432527/contributions/1072183/attachments/1321135/1982601/20160805_ICHEP_sato.pdf}{Talk presented at ICHEP 2016.}


  
  
  
  
    
\bibitem{Kiers:1997zt} 
  K.~Kiers and A.~Soni,
  Phys.\ Rev.\ D {\bf 56}, 5786 (1997)
  doi:10.1103/PhysRevD.56.5786
  [hep-ph/9706337].
  
\bibitem{Chen:2006nua} 
  C.~H.~Chen and C.~Q.~Geng,
  JHEP {\bf 0610}, 053 (2006)
  doi:10.1088/1126-6708/2006/10/053
  [hep-ph/0608166].
  
\bibitem{Kamenik:2008tj} 
  J.~F.~Kamenik and F.~Mescia,
  Phys.\ Rev.\ D {\bf 78}, 014003 (2008)
  doi:10.1103/PhysRevD.78.014003
  [arXiv:0802.3790 [hep-ph]].
  
\bibitem{Nierste:2008qe} 
  U.~Nierste, S.~Trine and S.~Westhoff,
  Phys.\ Rev.\ D {\bf 78}, 015006 (2008)
  doi:10.1103/PhysRevD.78.015006
  [arXiv:0801.4938 [hep-ph]].
  
\bibitem{Fajfer:2012vx} 
  S.~Fajfer, J.~F.~Kamenik and I.~Nisandzic,
  Phys.\ Rev.\ D {\bf 85}, 094025 (2012)
  doi:10.1103/PhysRevD.85.094025
  [arXiv:1203.2654 [hep-ph]].
  
\bibitem{Na:2015kha} 
  H.~Na {\it et al.} [HPQCD Collaboration],
  Phys.\ Rev.\ D {\bf 92}, no. 5, 054510 (2015)
  doi:10.1103/PhysRevD.92.054510
  [arXiv:1505.03925 [hep-lat]].
  
 
\bibitem{Bailey:2014tva} 
  J.~A.~Bailey {\it et al.} [Fermilab Lattice and MILC Collaborations],
  Phys.\ Rev.\ D {\bf 89}, no. 11, 114504 (2014)
  doi:10.1103/PhysRevD.89.114504
  [arXiv:1403.0635 [hep-lat]].

\bibitem{Lees:2013uzd} 
  J.~P.~Lees {\it et al.} [BaBar Collaboration],
  Phys.\ Rev.\ D {\bf 88}, no. 7, 072012 (2013)
  doi:10.1103/PhysRevD.88.072012
  [arXiv:1303.0571 [hep-ex]].
  
\bibitem{Huschle:2015rga} 
  M.~Huschle {\it et al.} [Belle Collaboration],
  Phys.\ Rev.\ D {\bf 92}, no. 7, 072014 (2015)
  doi:10.1103/PhysRevD.92.072014
  [arXiv:1507.03233 [hep-ex]].
  
\bibitem{Abdesselam:2016cgx} 
  A.~Abdesselam {\it et al.} [Belle Collaboration],
  arXiv:1603.06711 [hep-ex].
  
\bibitem{Abdesselam:2016xqt} 
  A.~Abdesselam {\it et al.},
  arXiv:1608.06391 [hep-ex].
    
\bibitem{Aaij:2015yra} 
  R.~Aaij {\it et al.} [LHCb Collaboration],
  Phys.\ Rev.\ Lett.\  {\bf 115}, no. 11, 111803 (2015)
  [Phys.\ Rev.\ Lett.\  {\bf 115}, no. 15, 159901 (2015)]
  doi:10.1103/PhysRevLett.115.159901, 10.1103/PhysRevLett.115.111803
  [arXiv:1506.08614 [hep-ex]].

\bibitem{Lees:2012ju} 
  J.~P.~Lees {\it et al.} [BaBar Collaboration],
  Phys.\ Rev.\ D {\bf 88}, no. 3, 031102 (2013)
  doi:10.1103/PhysRevD.88.031102
  [arXiv:1207.0698 [hep-ex]].



\bibitem{Kronenbitter:2015kls} 
  B.~Kronenbitter {\it et al.} [Belle Collaboration],
  Phys.\ Rev.\ D {\bf 92}, no. 5, 051102 (2015)
  doi:10.1103/PhysRevD.92.051102
  [arXiv:1503.05613 [hep-ex]].
  
  
\bibitem{Flynn:2015mha} 
  J.~M.~Flynn, T.~Izubuchi, T.~Kawanai, C.~Lehner, A.~Soni, R.~S.~Van de Water and O.~Witzel,
  Phys.\ Rev.\ D {\bf 91}, no. 7, 074510 (2015)
  doi:10.1103/PhysRevD.91.074510
  [arXiv:1501.05373 [hep-lat]],\, 

  
\bibitem{Lattice:2015tia} 
  J.~A.~Bailey {\it et al.} [Fermilab Lattice and MILC Collaborations],
  Phys.\ Rev.\ D {\bf 92}, no. 1, 014024 (2015)
  doi:10.1103/PhysRevD.92.014024
  [arXiv:1503.07839 [hep-lat]].
  
\bibitem{Detmold:2015aaa}
  W.~Detmold, C.~Lehner and S.~Meinel,
  Phys.\ Rev.\ D {\bf 92}, no. 3, 034503 (2015)
  doi:10.1103/PhysRevD.92.034503
  [arXiv:1503.01421 [hep-lat]].

\bibitem{Na:2012kp} 
  H.~Na, C.~J.~Monahan, C.~T.~H.~Davies, R.~Horgan, G.~P.~Lepage and J.~Shigemitsu,
  Phys.\ Rev.\ D {\bf 86}, 034506 (2012)
  doi:10.1103/PhysRevD.86.034506
  [arXiv:1202.4914 [hep-lat]].
  
\bibitem{Aoki:2013ldr} 
  S.~Aoki {\it et al.},
  Eur.\ Phys.\ J.\ C {\bf 74}, 2890 (2014)
  doi:10.1140/epjc/s10052-014-2890-7
  [arXiv:1310.8555 [hep-lat]].

  \bibitem{GF}
  \url{http://pdg.lbl.gov/2015/reviews/rpp2014-rev-phys-constants.pdf}
 
\bibitem{mB}
  \url{http://pdglive.lbl.gov/DataBlock.action?node=S041M}
  
\bibitem{mtau}
  \url{http://pdglive.lbl.gov/DataBlock.action?node=S035M}
  
\bibitem{tauB+}
  \url{http://pdg.lbl.gov/2015/tables/rpp2015-tab-mesons-bottom.pdf}
  
  
  
\bibitem{Amsler:2008zzb} 
  C.~Amsler {\it et al.} [Particle Data Group Collaboration],
  Phys.\ Lett.\ B {\bf 667}, 1 (2008).
  doi:10.1016/j.physletb.2008.07.018
  
\bibitem{Alberti:2014yda} 
  A.~Alberti, P.~Gambino, K.~J.~Healey and S.~Nandi,
  Phys.\ Rev.\ Lett.\  {\bf 114}, no. 6, 061802 (2015)
  doi:10.1103/PhysRevLett.114.061802
  [arXiv:1411.6560 [hep-ph]].
  
\bibitem{Xing:2007fb} 
  Z.~z.~Xing, H.~Zhang and S.~Zhou,
  Phys.\ Rev.\ D {\bf 77}, 113016 (2008)
  doi:10.1103/PhysRevD.77.113016
  [arXiv:0712.1419 [hep-ph]].
  
\bibitem{Carrasco:2014cwa} 
  N.~Carrasco {\it et al.} [European Twisted Mass Collaboration],
  Nucl.\ Phys.\ B {\bf 887}, 19 (2014)
  doi:10.1016/j.nuclphysb.2014.07.025
  [arXiv:1403.4504 [hep-lat]].
  
\bibitem{Grossman:1994ax} 
  Y.~Grossman and Z.~Ligeti,
  Phys.\ Lett.\ B {\bf 332}, 373 (1994)
  doi:10.1016/0370-2693(94)91267-X
  [hep-ph/9403376, hep-ph/9403376].
  
\bibitem{Bigi:2016mdz} 
  D.~Bigi and P.~Gambino,
  arXiv:1606.08030 [hep-ph].
  
\bibitem{Lattice:2015rga} 
  J.~A.~Bailey {\it et al.} [MILC Collaboration],
  Phys.\ Rev.\ D {\bf 92}, no. 3, 034506 (2015)
  doi:10.1103/PhysRevD.92.034506
  [arXiv:1503.07237 [hep-lat]].


  
  
  
  

  

  

  

  
  
  \bibitem{Falk:1994gw} 
  A.~F.~Falk, Z.~Ligeti, M.~Neubert and Y.~Nir,
  Phys.\ Lett.\ B {\bf 326}, 145 (1994)
  doi:10.1016/0370-2693(94)91206-8
  [hep-ph/9401226].

\bibitem{Ligeti:2014kia} 
  Z.~Ligeti and F.~J.~Tackmann,
  Phys.\ Rev.\ D {\bf 90}, no. 3, 034021 (2014)
  doi:10.1103/PhysRevD.90.034021
  [arXiv:1406.7013 [hep-ph]].


\bibitem{Bernlochner:2012bc} 
  F.~U.~Bernlochner, Z.~Ligeti and S.~Turczyk,
  Phys.\ Rev.\ D {\bf 85}, 094033 (2012)
  doi:10.1103/PhysRevD.85.094033
  [arXiv:1202.1834 [hep-ph]].

\bibitem{Amhis:2012bh} 
  Y.~Amhis {\it et al.} [Heavy Flavor Averaging Group Collaboration],
  arXiv:1207.1158 [hep-ex].

  
\bibitem{Beringer:1900zz} 
  J.~Beringer {\it et al.} [Particle Data Group Collaboration],
  Phys.\ Rev.\ D {\bf 86}, 010001 (2012).
  doi:10.1103/PhysRevD.86.010001
  
  
  
\bibitem{Tanaka:1994ay} 
  M.~Tanaka,
  Z.\ Phys.\ C {\bf 67}, 321 (1995)
  doi:10.1007/BF01571294
  [hep-ph/9411405].


 \bibitem{Kim:2007uq} 
  C.~S.~Kim and R.~M.~Wang,
  Phys.\ Rev.\ D {\bf 77}, 094006 (2008)
  doi:10.1103/PhysRevD.77.094006
  [arXiv:0712.2954 [hep-ph]].

\bibitem{Khodjamirian:2011ub} 
  A.~Khodjamirian, T.~Mannel, N.~Offen and Y.-M.~Wang,
  Phys.\ Rev.\ D {\bf 83}, 094031 (2011)
  doi:10.1103/PhysRevD.83.094031
  [arXiv:1103.2655 [hep-ph]].

  
  
\bibitem{Bernlochner:2015mya} 
  F.~U.~Bernlochner,
  Phys.\ Rev.\ D {\bf 92}, no. 11, 115019 (2015)
  doi:10.1103/PhysRevD.92.115019
  [arXiv:1509.06938 [hep-ph]].

  \bibitem{Bellevub}
  A.~Sibidanov {\it et al.} [Belle Collaboration],
  Phys.\ Rev.\ D {\bf 88}, no. 3, 032005 (2013)
  doi:10.1103/PhysRevD.88.032005
  [arXiv:1306.2781 [hep-ex]].

  \bibitem{Babarvub}
  B.~Aubert {\it et al.} [BaBar Collaboration],
  Phys.\ Rev.\ Lett.\  {\bf 101}, 081801 (2008)
  doi:10.1103/PhysRevLett.101.081801
  [arXiv:0805.2408 [hep-ex]].

\bibitem{Du:2015tda} 
  D.~Du, A.~X.~El-Khadra, S.~Gottlieb, A.~S.~Kronfeld, J.~Laiho, E.~Lunghi, R.~S.~Van de Water and R.~Zhou,
  Phys.\ Rev.\ D {\bf 93}, no. 3, 034005 (2016)
  doi:10.1103/PhysRevD.93.034005
  [arXiv:1510.02349 [hep-ph]].
  
\bibitem{Hamer:2015jsa} 
  P.~Hamer {\it et al.} [Belle Collaboration],
  Phys.\ Rev.\ D {\bf 93}, no. 3, 032007 (2016)
  doi:10.1103/PhysRevD.93.032007
  [arXiv:1509.06521 [hep-ex]].
  
\bibitem{Agashe:2014kda} 
  K.~A.~Olive {\it et al.} [Particle Data Group Collaboration],
  Chin.\ Phys.\ C {\bf 38}, 090001 (2014).
  doi:10.1088/1674-1137/38/9/090001

  \bibitem{Glashow:1976nt} 
  S.~L.~Glashow and S.~Weinberg,
  Phys.\ Rev.\ D {\bf 15}, 1958 (1977).
  doi:10.1103/PhysRevD.15.1958

  \bibitem{Paschos:1976ay} 
  E.~A.~Paschos,
  Phys.\ Rev.\ D {\bf 15}, 1966 (1977).
  doi:10.1103/PhysRevD.15.1966
  
\bibitem{Hou:1992sy} 
  W.~S.~Hou,
  Phys.\ Rev.\ D {\bf 48}, 2342 (1993).
  doi:10.1103/PhysRevD.48.2342
  
\bibitem{Akeroyd:2007eh} 
  A.~G.~Akeroyd and C.~H.~Chen,
  Phys.\ Rev.\ D {\bf 75}, 075004 (2007)
  doi:10.1103/PhysRevD.75.075004
  [hep-ph/0701078].
  
\bibitem{Akeroyd:2003zr} 
  A.~G.~Akeroyd and S.~Recksiegel,
  J.\ Phys.\ G {\bf 29}, 2311 (2003)
  doi:10.1088/0954-3899/29/10/301
  [hep-ph/0306037].
  
\bibitem{Akeroyd:2003jb} 
  A.~G.~Akeroyd,
  Prog.\ Theor.\ Phys.\  {\bf 111}, 295 (2004)
  doi:10.1143/PTP.111.295
  [hep-ph/0308260].
  
 
\bibitem{Akeroyd:2009tn} 
  A.~G.~Akeroyd and F.~Mahmoudi,
  JHEP {\bf 0904}, 121 (2009)
  doi:10.1088/1126-6708/2009/04/121
  [arXiv:0902.2393 [hep-ph]].

\bibitem{Crivellin:2012ye} 
  A.~Crivellin, C.~Greub and A.~Kokulu,
  Phys.\ Rev.\ D {\bf 86}, 054014 (2012)
  doi:10.1103/PhysRevD.86.054014
  [arXiv:1206.2634 [hep-ph]].
  
  
  
\bibitem{Sakaki:2014sea} 
  Y.~Sakaki, M.~Tanaka, A.~Tayduganov and R.~Watanabe,
  Phys.\ Rev.\ D {\bf 91}, no. 11, 114028 (2015)
  doi:10.1103/PhysRevD.91.114028
  [arXiv:1412.3761 [hep-ph]].
  
  
 
\bibitem{Pesantez:2015qjc} 
  L.~Pes$\acute{a}$ntez,
  PoS FPCP {\bf 2015}, 012 (2015).
  \url{http://pos.sissa.it/archive/conferences/248/012/FPCP2015_012.pdf}
  
\bibitem{Lunghi:2009ke} 
  E.~Lunghi and A.~Soni,
  Phys.\ Rev.\ Lett.\  {\bf 104}, 251802 (2010)
  doi:10.1103/PhysRevLett.104.251802
  [arXiv:0912.0002 [hep-ph]].

\end{thebibliography}
\end{document}